\documentclass[12pt]{article}
\usepackage{amssymb, amsmath ,amsthm , bm, float, epsfig, color}
\usepackage[centering]{geometry}
\usepackage{braket}
\usepackage{slashed}
\usepackage{multirow}
\usepackage{multicol}
\usepackage{array}
\usepackage{comment}

\newcommand{\p}{\partial}
\newcommand{\mb}[1]{{\mathbb{#1}}}
\newcommand{\mc}[1]{{\mathcal{#1}}}

\newcolumntype{C}[1]{>{\centering}p{#1\linewidth}}
\newcommand{\vs}{\vspace{0.3cm}}

\theoremstyle{definition}

\newtheorem*{Def*}{Definition}

\newcommand{\acom}[1]{{\left\{#1\right\}}}
\newcommand{\fl}[1]{{\left\lfloor #1 \right\rfloor}}

\makeatletter
\@addtoreset{equation}{section}
\makeatother

\begin{document}
\title{
\vspace{-30pt}
\begin{flushright}
\normalsize WU-HEP-19-03 \\
DESY 19-018 \\*[55pt]
\end{flushright}
{\Large\bf Wavefunctions on $S^2$ with flux and branes\\*[20pt]}
}

\author{
Sosuke~Imai,$^1$\footnote{E-mail address: s.i.sosuke@akane.waseda.jp} \quad and \quad
Yoshiyuki~Tatsuta,$^{2}$\footnote{E-mail address: yoshiyuki.tatsuta@desy.de}\\*[30pt]
$^1${\it \normalsize Department of Physics, Waseda University, Tokyo 169-8555, Japan}\\
$^2${\it \normalsize Deutsches Elektronen-Synchrotron DESY, Hamburg 22607, Germany}\\*[55pt]
}

\date{
\centerline{\small \bf Abstract}
\begin{minipage}{0.9\textwidth}
\medskip\medskip
\small
We formulate a six dimensional $U(1)$ gauge theory compactified on a (two dimensional) sphere $S^2$ with flux and localized brane sources.
Profiles of the lowest Kaluza--Klein (KK) wavefunctions and their masses are derived analytically.
In contrast to ordinary sphere compactifications, the above setup can lead to the degeneracy of and the sharp localizations of the linearly independent lowest KK modes, depending on the number of branes and their tensions.
Moreover, it can naturally accommodate CP violation in Yukawa interactions.
\end{minipage}}

\begin{titlepage}
\maketitle
\thispagestyle{empty}
\end{titlepage}
\tableofcontents

\section{Introduction}
Flux compactifications play an important role in phenomenological and cosmological models based on (super)string theories as well as higher dimensional theories.
These may explain underlying structures in the Standard Model (SM) of particle physics, such as chiral matters, multiplicities in the SM fermions, Yukawa couplings, the wine-bottle potential of the Higgs boson and so forth.
Often, higher dimensional theories are non-chiral, so a mechanism has to be invoked to produce chirality in four-dimensional (4d) spacetime.
One powerful way to {achieve this} is a flux compactification.
Supposing that higher dimensional space is compactified on some compact manifold $X$ with a non-vanishing flux background, the vector-like matter is reduced to a chiral matter in the low energy theory.
Also, Kaluza--Klein (KK) zero modes of chiral matters are multiply degenerated due to the index theorem \cite{Atiyah:1963zz,Cremades:2004wa,Conlon:2008qi}.
In the low energy effective theory of such models, it is justified to identify degenerated zero modes as family multiplicities.
Various phenomenological models of particle physics and cosmology have been constructed by use of such a ``family generating'' mechanism \cite{Abe:2008sx,Abe:2012fj,Abe:2013bba,Abe:2015yva,Buchmuller:2015jna}.

Once the zero mode wavefunctions of a chiral family are known, it is straightforward to calculate coupling constants of the low energy effective theory, in particular, Yukawa couplings by integrating
\begin{gather}
\label{YUKAWA}
y_{ijk} \sim g \int_{X} (\psi^i)^\dag \psi^j \phi^k \, {\rm vol}(X),
\end{gather}
where $g$ is a coupling constant of the higher dimensional theory.
The Yukawa couplings are given by overlap integrals of zero modes.
Examples are the two-dimensional (2d) torus \cite{Cremades:2004wa}, the 2d (pure) sphere \cite{Conlon:2008qi} or higher genus Riemann surfaces.

A particularly interesting case is that zero mode wavefunctions of three {species} of matter are localized at different positions in the manifold.
In such a situation, as proposed by Arkani-Hamed and Schmaltz \cite{ArkaniHamed:1999dc} in 1999 and in subsequent investigations \cite{Kaplan:2000av, Huber:2000ie}, this can lead to considerably small elements in the Yukawa matrix.
This would explain the hierarchical structure of the SM Yukawa matrices from a simple higher dimensional point of view.
For those reasons, flux compactifications have received much attention in the context of string phenomenology as well as in usual bottom-up model building.
Even today, it is one of the key issues to solve the flavor problem using extra dimensions.
Indeed, numerous approaches were followed, for example, \cite{Abe:2014vza,Fujimoto:2016zjs,Kobayashi:2016qag}.
As well as the family generating mechanism, the (bulk) Yukawa couplings have been utilized to construct phenomenological models, for example, non-abelian discrete flavor symmetry in leptons \cite{Abe:2009vi,Abe:2014nla} and for inflation \cite{Higaki:2016ydn}.

A previous study \cite{Conlon:2008qi}, carried out by Conlon, Maharana and Quevedo in 2008, marvelously described a setup of flux compactification on a two dimensional sphere $S^2$, almost 30 years after Wu and Yang originally proposed a magnetic monopole on $S^2$ \cite{Wu:1976ge, Wu:1976qk}.
The two dimensional sphere is one of the simplest manifolds with known higher dimensional metric and non-trivial curvature.
One of the main motivations in Ref.\,\cite{Conlon:2008qi} was to explore and formulate sphere compactifications with non zero flux as local models in string compactifications.
In contrast to global models, which are for example compactified on a torus or toroidal orbifolds, local models only reflect the property of singularities independently of bulk properties.
In particular, it is concretely described that the profile of KK zero mode wavefunctions is determined only by properties of singularity.
Also, since the sphere metric is treated as a solution of the gravitational equation (the Liouville equation), we can take into account backreaction in the presence of flux and branes.
Thus, the sphere compactification with non zero flux and curvature has been expected to be a promising setup for phenomenological model building which can give rise to family replication and realistic flavor patterns.
Moreover, it is clearly described that massless zero modes of $U(1)$ charged matter can be expressed analytically after an appropriate $U(1)$ charge quantization is imposed, that is, the Dirac charge quantization.
However, using the analytic expressions of zero modes, it was also argued that Yukawa coupling elements derived from overlapping wavefunctions are not hierarchical enough in order to explain the observed hierarchies in the quark and charged lepton sectors.

In this paper, we reformulate a $U(1)$ gauge theory on $S^2$ with magnetic flux background in the presence of geometrical singularities (called ``branes'' in this paper).
The $S^2$ metric $G$ obeys the Liouville equation
\begin{gather}
	{-\frac{4}{G^2}\p\bar\partial \log G=k+2\pi\sum_{a=1}^{N}\alpha_a \delta^2 (z -z_a)},
\end{gather}
where $N$ denotes the number of branes, localized at positions $z_a$ and tensions $\alpha_a$.
This can be solved analytically up to $N=3$ \cite{Troyanov:1989,Umehara:1998,Redi:2004tm,Eremenko:2004}.
By use of such results, we demonstrate that forms of KK zero mode wavefunctions are strongly peaked in the vicinity of the brane positions, argued in Ref.\,\cite{Conlon:2008qi}.
In addition, we analytically derive zero mode wavefunctions and comment on their Yukawa couplings via overlap integrals.
Thus, the reformulation of $S^2$ with flux and branes is well motivated as the first step to explore (realistic) model building.
The formulation in this paper is expected to reveal potential possibilities of sphere compactification and also to extend to more general manifolds or orbifolds possessing more complicated geometries.

This paper is organized as follows.
In the next section, we review the formulation of KK wavefunctions in {Ref.\,}\cite{Conlon:2008qi}.
In Sec.\,3, we derive analytically KK zero mode wavefunctions in the presence of multiple branes on the 2d sphere, and then depict profiles of wavefunctions, {and discuss  modifications of KK masses due to branes.}
In Sec.\,4, an example of Yukawa couplings is presented and we explain a strategy to realize (semi-)realistic flavor patterns.
Section\,5 is devoted to conclusion and discussion.
In Appx.\,A, we explain patches and gauge transformations on overlaps in the flux compactification on $S^2$, following \cite{Conlon:2008qi}.
In Appx.\,B, we show a way to count the number of zero modes, based on the index theorem.
In Appx.\,C, we comment on the isometries of the sphere with flux and branes.

\section{Short review of wavefunctions and Yukawa couplings on $S^2$ with flux}
{Higher dimensional gauge theories} compactified on projective spaces $\mb{CP}^1 \,\, (\simeq S^2)$, $\mb{CP}^1 \times \mb{CP}^1$ {and} $\mb{CP}^2$, with (abelian) magnetic flux are formulated in Ref.\,\cite{Conlon:2008qi}.
The authors {analytically derived zero mode} wavefunctions and {calculated} Yukawa couplings {given by overlap integrals as} Eq.\,\eqref{YUKAWA}.
This section is devoted to providing a short review for the case of 2d  sphere $S^2 = \mb{C}\cup \acom{\infty}$.\,\footnote{
Note that 2d sphere $S^2$ is diffeomorphic to the complex plane $\mathbb{C}$ once a point on $S^2$ is removed, and $\mathbb{C} \cup \{\infty\}$ is also called the Riemann sphere.
}

The Fubini--Study metric of $S^2$ {reads}
\begin{gather}
	ds^2 = \frac{2}{k(1+|z|^2)^2}(dz\otimes d\bar{z}+d\bar{z}\otimes dz)
\end{gather}
with a constant Gaussian curvature $k>0$.
Here, $z$ is a complex coordinate parameterizing the {``south hemisphere'' of $S^2$ and then $z \in \mb{C} = S^2 \setminus \acom{\infty}$}.
The northern patch takes the same form, with $w = - 1/z$ on the overlap between patches.
Lorentz invariance in 4d prohibits non-vanishing vacuum expectation values (VEVs) of gauge fields along 4d spacetime.
{On the other hand, it is still} possible for gauge fields to have non-vanishing VEVs along extra dimensions $z$ and $\bar z$.
When an abelian gauge field $A$ {develops its VEV}, its field strength $F=dA$ {as flux background should satisfy the Maxwell equation} and $F$ reads
\begin{gather}
	F=\frac{iM}{(1+|z|^2)^2}dz\wedge d\bar{z},
\end{gather}
where $M$ {denotes a magnitude of magnetic flux and should be quantized appropriately as explained later and in Appx.\,A.
We call $M$ ``flux'' in what follows.}
{The covariant derivative $D=d-\omega E-iAQ$ on $S^2$ with flux background is characterized by the spin connection $\omega$ and the vector potential $A$ associated with the flux background $F$}, where $E$ is {a} generator of $SO(2)$ along the extra dimensions, $Q$ is {a} generator of $U(1)$ gauge symmetry and $d$ is an ordinary derivative {with respect to $z$ and $\bar z$}.

\begin{table}[t]
\centering

	\begin{tabular}{|cc|C{0.1}|C{0.12}|c|C{0.12}|c|}\hline
	                                            \multicolumn{2}{
|c|}{}                                                                                       &    \multirow{2}{*}{$s$}       &                                   \multicolumn{2}{c|}{$qM>0$}                              &        \multicolumn{2}{c|}{$qM<0$}         \\  \cline{4-7}\rule[-7pt]{0pt}{21pt}
		                                          &                                                                                                              &                                         &      $i_{\rm max}$       &    \hspace{0.3cm}$m_{\rm KK}^2$\hspace{0.3cm} &          $i_{\rm max}$       &       \hspace{0.3cm}$m_{\rm KK}^2$\hspace{0.3cm}    \\ \hline\hline \rule[-7pt]{0pt}{21pt}
		\multirow{2}{*}{Scalar}       &                             $\phi\,\,\,(\bar D\phi=0)$                                      &        $\frac{qM}{2}$	     &         $|qM|+1$            &    $\frac{k|qM|}{2}$                                          &                        ---              &           ---                                          \\ \rule[-7pt]{0pt}{21pt}
		                                           &                               $\phi\,\,\,(D\phi=0)$                                  &        $\frac{-qM}{2}$	     &         ---                           &             ---                                                        &           $|qM|+1$              &           $\frac{k|qM|}{2}$                                           \\ \hline\rule[-7pt]{0pt}{21pt}

		\multirow{2}{*}{Spinor}       &  \multirow{2}{*}{$\psi=\begin{pmatrix} \psi_- \\ \psi_+ \end{pmatrix} $} &        $\frac{qM-1}{2}$	     &         $|qM|$                &                 $0$                                                  &           ---                            &          ---                                                                    \\ \rule[-7pt]{0pt}{21pt}
		                                          &                                                                                                              &        $\frac{-qM-1}{2}$     &          ---                          &                           ---                                              &           $|qM|$                 &          $0$                                                                 \\ \hline\rule[-7pt]{0pt}{21pt}
		\multirow{2}{*}{Vector}       &                                     $B_z$                                                               &        $\frac{qM}{2}$         &         $|qM|-1$             &    $-\frac{k|qM|}{2}$                                         &           ---                            &         ---                                                                 \\ \rule[-7pt]{0pt}{21pt}
		                                          &                                     $B_{\bar z}$                                                     &        $\frac{-qM}{2}$         &          ---                       &              ---                                                        &           $|qM|-1$              &       $-\frac{k|qM|}{2}$ \\ \hline
	\end{tabular}
	\caption{
		The power $s$, the degeneracy $i_{\rm max}$ and the {KK mass $m_{\rm KK}^2$ for the lowest wavefunctions whose $U(1)$ charge is $q$ (see \cite{Conlon:2008qi} for more details).}
	}
\end{table}

Zero mode wavefunctions of the {6d Weyl} spinor field are obtained by solving the massless Dirac equation on $S^2$.\,\footnote{
The Pauli matrices $\sigma^i\,\,(i=1,2)$ satisfy the 2d Clifford algebra $\{\sigma^i,\sigma^j\}=2\delta^{ij}$.
To define 6d spinor fields, as 6d gamma matrices, we use tensor products of the Pauli matrices and 4d gamma matrices $\gamma^\mu$ satisfying $\{\gamma^\mu,\,\gamma^\nu\}=2\eta^{\mu\nu}$.
}
{For $U(1)$ charge $Q=q$ of $\psi$}, the Dirac operator $\slashed{D}=\sigma^iD_i$ {in terms of the Pauli matrices} $\sigma^i \,\, (i=1,2)$ is given as
\begin{gather}
	\slashed{D}\propto
\begin{pmatrix} & D \\ \bar D & \end{pmatrix} =
\begin{pmatrix} & \p+\frac{\bar{z}(-qM-1)}{2(1+|z|^2)} \\ \bar{\partial}+\frac{z(qM-1)}{2(1+|z|^2)} & \end{pmatrix}
\end{gather}
up to an overall {(positive)} function.
Here, $\p$ and $\bar \partial$ are derivatives with respect to $z$ and $\bar{z}$, respectively.
The general solution of {the corresponding} zero mode equation $\slashed{D}\psi=0$ is {given by}
\begin{gather}
	\psi=\begin{pmatrix}(1+|z|^2)^{\frac{1-qM}{2}}\,u_- \\ (1+|z|^2)^{\frac{1+qM}{2}}\,\bar{u}_+ \end{pmatrix},
\end{gather}
where $u_{\mp}$ are holomorphic functions {of $z$} on $\mb{C}\subseteq S^2$.
Since $\psi$ must be well-defined and regular on the northern patch, {the Dirac quantization condition for flux $qM\in \mb{Z}$ is necessary in terms of two patches covering the south and north hemispheres of $S^2$, as reviewed in Appx.\,A.}
In addition, to {lead to} a physically meaningful low energy effective theory, the wavefunction $\psi$ must be normalizable, so {the norm of $\psi$ must be finite,}
\begin{gather}
	\braket{\psi|\psi} {\equiv} \int_{S^2}\bar{\psi}\psi\,{\rm vol}(S^2)<\infty,
\end{gather}
{otherwise, the kinetic term in the low energy effective theory diverges.}
{Due to this normalizability of wavefunctions, the vector space spanned by the solutions is required to be finite dimensional}, and we obtain the following {as an orthonormal basis}
\begin{gather}\begin{array}{rccl}
	qM>0:& & \qquad \psi^i = \mc{N}_i(1+|z|^2)^{\frac{1-qM}{2}} \begin{pmatrix} z^{i-1} \\ 0 \end{pmatrix} & (i=1,\dots, {qM}), \vspace{0.3cm} \\
	qM<0:& & \qquad \psi^i = \mc{N}_i(1+|z|^2)^{\frac{1+qM}{2}}\begin{pmatrix} 0 \\  \bar{z}^{i-1} \end{pmatrix} & (i=1,\dots,|qM|).
\end{array}\end{gather}
The normalization factor $\mc{N}_i \,\, (>0)$ is determined {so that} $\braket{\psi^i|\psi^i}=1$.
When the spinor field is neutral {under $U(1)$ associated with flux, i.e.,} $Q=0$, {there is} no zero mode because of the positive curvature of $S^2$.
{It can be observed} that there are only {zero modes for 4d} Weyl or anti-Weyl depending on the sign of $qM$.
This can be {considered} as {an} origin of the chiral spectrum {and we can relate the degeneracy $|qM|$ of zero modes to the family structure in the SM fermions.}

Similar results are obtained for a complex scalar $\phi$ and a complex vector (1-form) $B=B_zdz+B_{\bar{z}}d\bar{z}$.
All of {the zero mode wavefunctions} on the {magnetized $S^2$} consist of {a} metric factor $(1+|z|^2)^{-s}$ and a holomorphic (or anti-holomorphic) part $z^{i-1}$ (or $\bar{z}^{i-1}$):
\begin{gather}
	\phi,\,\psi_{\mp},\,B_{z,\bar{z}}\hspace{0.3cm}\propto\hspace{0.3cm}(1+|z|^2)^{-s}z^{i-1}\hspace{0.3cm}{\rm or}\hspace{0.3cm}(1+|z|^2)^{-s}\bar{z}^{i-1}\hspace{0.5cm}(i=1,\dots,i_{\rm max}),
\end{gather}
where $\psi_{\mp}$ are components of the {2d} spinor field $\psi=(\psi_-,\,\psi_+)$.
The power $s$ and the degeneracy $i_{\rm max}$ are determined by the spin {of fields} and $U(1)$ charge {$q$}.
{The holomorphic part $z^{i-1}$ (the anti-holomorphic part $\bar{z}^{i-1}$) appears when $qM>0$ ($qM<0$).}
There {is no zero mode} for $\psi_+$ and $B_{\bar{z}}$ when $qM\geq0$ and, while there {is no zero mode} for $\psi_-$ and $B_z$ when $qM\leq0$ due to the {normalizability of zero modes}.
{For the complex scalar field,} we have two possibilities for the zero mode equation, $D\phi=0$ or $\bar D \phi =0$, but non-trivial solutions exist only for one of them if $q M \neq 0$.
Note that the lowest KK mass is not always zero for scalars and vectors once the flux is introduced (see Table.\,1).\,\footnote{{Such situations can be seen also on the magnetized torus \cite{Cremades:2004wa}, for instance.}}
The power $s$, the degeneracy $i_{\rm max}$ and the {KK} mass $m_{\rm KK}^2$ of the lowest wavefunctions for each representation are summarized in Tab.\,1.
{In Ref.\,\cite{Conlon:2008qi}, the complex vector field $B$ is considered as a part of a higher dimensional Yang--Mills (YM) field based on a compact Lie group.
When the Cartan directions of the YM field develops their VEVs, the kinetic term of the YM field contains a quadratic coupling of $B$ proportional to the abelian flux, and the tachyonic KK mass $- k|qM|/2$ of the complex vector $B$ comes from the quadratic coupling.
In this paper, we also suppose that $B$ is a part of some YM field, and the KK mass originating from the YM kinetic term is taken into consideration. }

\begin{figure}[t]
\centering
\includegraphics[width=0.6\textwidth]{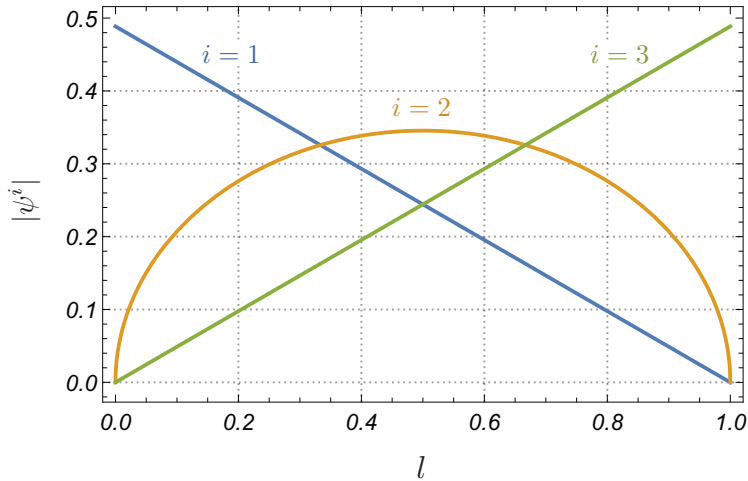}
\caption{Profiles of absolute values of spinor wavefunctions in $l$-coordinate with $k=1$ and $qM=3$}
\end{figure}

At the end of this section, we qualitatively outline a hierarchy of Yukawa couplings given in Eq.\,\eqref{YUKAWA} which are proportional to overlap integrals
\begin{gather}\label{OVERLAP}
{y_{ijk} \propto} \int_0^\infty rdr\int_0^{2\pi}d\theta \, \frac{4}{k(1+r^2)^2}(\psi^i)^\dagger\psi^j \phi^k,
\end{gather}
where $\psi^i$ and $\psi^j$ are zero mode wavefunctions of spinor fields {such as the SM fermions} and $\phi^k$ is the lowest mode wavefunction of a boson {such as the Higgs boson}.
The integration is performed in a polar coordinate $z=re^{i\theta}\in \mb{C}\subseteq S^2$, and an additional weight {factor} $4/(k(1+r^2)^2)$ comes from the volume form of $S^2$.
To evaluate the integration {above}, it is useful to introduce a coordinate
\begin{gather}
	l=1-\frac{1}{1+r^2},
\end{gather}
{so that $l=0, 1/2$ and $1$ correspond to the {``south pole''}, the ``equator'' and the {``north pole''}, respectively.}
{It can be seen that} $dl$ absorbs the weight {factor} and the non-vanishing overlap integral \eqref{OVERLAP} reduces to
\begin{gather}
	{y_{ijk}\propto}\left.\frac{4\pi}{k}\int_0^1dl\, |\psi^i\psi^j \phi^k| \, \right|_{\theta=0}.
\end{gather}
{Wavefunction profiles} of $\psi^i$ with $|qM|=3$, which {would correspond to} three generation matters in the SM, are {depicted} in Fig.\,1.
This figure shows that $|\psi^i|$ spreads gently and entirely on $S^2$ and {accordingly the overlap integral \eqref{OVERLAP} tends to give sizable values not suitable for large hierarchies among the SM quarks and charged leptons.}\,\footnote{{The same profiles appear in the bosonic wavefunctions also and spread gently and entirely on $S^2$ in a similar manner.}}
Therefore, {it is fair to say that unless the profiles of bosonic wavefunctions $\phi^k$ are strongly localized,} the Yukawa couplings can not have a sufficiently hierarchical structure to reproduce the SM {flavor pattern}.

\section{Zero mode wavefunctions on $S^2$ with flux and branes}
\label{Wavefunctions}
\subsection{Two branes}
\label{TWOBRANES}
In this {subsection}, we {derive} zero mode wavefunctions in the presence of {additional singularities on $S^2$ such as orbifold fixed points}.
We call such an additional singularity a brane in the following.
Branes are supposed to be {4d} objects, fill {up the 4d spacetime} and couple only to the gravity.
{Then, the action reads}
\begin{gather}
S \equiv	S_{\rm matter}+S_{\rm gravity}+S_{\rm branes},
\end{gather}
{where} $S_{\rm matter}$ describes {6d} matter fields' dynamics and $S_{\rm gravity}$ is the Einstein--Hilbert action with a {6d} cosmological constant.
Although some discussions in the cosmological constant have been done in Refs.\,\cite{Salam:1984cj, Aghababaie:2003wz}, we do not touch the issue in this paper.
The last term describing field theories on {$N$} branes is {given as}
\begin{gather}
	S_{\rm branes}=-2\pi M_6^4\sum_{a=1}^N\alpha_a\int_{\mb{R}^{1,3}\times \acom{z_a}}{\rm vol}(\mb{R}^{1,3}\times \{z_a\}),
\end{gather}
where $M_6$ is the {6d Planck mass, $z_a \,\,(\in S^2)$ denotes a position at which} the $a$-th brane locates and a real constant $\alpha_a$ is {a} dimensionless brane tension {which the $a$-th brane possesses.}
Since {the dimensionless tension can restore its mass dimension as  $M_6^4\alpha_a$}, $\alpha_a$ is {assumed to be} positive because negative tension branes {would} be physically unstable.
We also assume that $\alpha_a<1$ because the conical singularity induced by the $a$-th brane has deficit angle $2\pi \alpha_a<2\pi$.
In this subsection, we focus on a situation with two branes ($N=2$) totally, where branes are located at $z_1=0$ and $z_2=\infty$.\,\footnote{
Two dimensional sphere has $SL(2,\mb{C})$ symmetry. This fixes two branes on $z=0,\,\infty$ without loss of generality.}
{Thus}, we treat the following {setup}\,:
\begin{gather}
\begin{array}{rcl}
	ds^2&=&\displaystyle\eta_{\mu\nu}dx^\mu\otimes dx^\nu+\frac{1}{2}G(z)^2(dz\otimes d\bar{z}+d\bar{z}\otimes dz), \vspace{0.3cm}\\
	F^i&=&\displaystyle\frac{i}{2}f^i G{(z)^2}dz\wedge d\bar{z},
\end{array}
\end{gather}
where $F^i$ is an {extra {dimensional} component of an abelian field strength}, or more generally a Cartan direction of non-abelian {field strength}, a real constant $f^i$ {is a flux density of flux background} and $\eta_{\mu\nu}$ is the {4d} Minkowski metric.
{The second equation is a solution of the Maxwell equation in the vacuum.
By substituting {these expressions} into the Einstein equation, we {obtain} the Liouville equation {which the extra dimensional metric $G$ should satisfy}\,:
\begin{gather}
	-\frac{4}{G^2}\p\bar\partial \log G=k+2\pi\sum_{a=1}^2\alpha_a {\delta^2 (z -z_a).}
\end{gather}
A real constant $k$ {denotes a curvature characterized by the flux density $f^i$ and {the 6d cosmological constant}.}
{In the whole of this paper,} we assume that $k$ is positive.
Also, $\delta^2(z - z_a)$ is a delta function on $S^2$ normalized as $\int_{S^2}\delta^2 {(z - z_a)} \varphi{(z)} \, {\rm vol}(S^2)=\varphi(z_a)$.
{From a geometrical point of view}, the left hand side {corresponds to} the Gaussian curvature of $G^2 dzd\bar z$.
Thus, the Liouville equation {above} tells that the Gaussian curvature of {this setup} is constant except on brane positions {$z = z_a$}.
The solution of the Liouville equation is investigated in Refs.\,\cite{Troyanov:1991,Redi:2004tm}.
It {exists} if $\alpha_1=\alpha_2=\alpha$, {given as}
\begin{gather}
\label{TWOBRANESBACKGROUND}
	G=\frac{1-\alpha}{\sqrt k}\frac{2|z|^{-\alpha}}{1+|z|^{2-2\alpha}}.
\end{gather}
{This metric diverges at $z=z_1=0$ and it can be observed that the metric also  diverges at $z=z_2=\infty$ after changing a patch with $w= -1/z$ (see Appx.\,A).}
These {metric singularities are called conical singularities.
 They} can be seemingly removed by introducing a {(singular)} new {coordinate} $z'=z^{1-\alpha}$.
In the {coordinate}, {there is no region of} $\{z'\in \mb{C}\mid 2\pi(1-\alpha)\leq {\rm arg}\,z'<2\pi\}$.
Schematically {the lack of the region leads to deficit angle $2\pi \alpha$ from a full domain of} $S^2$.
Thus {a shape of the sphere with two branes seems} ``football", {as depicted in Fig.\,2}.
Note that a constant curvature $k$ on $S^2$ with two branes is given as
\begin{gather}
k = 1 - \alpha
\end{gather}
when the volume of sphere is normalized as $\int{\rm vol}(S^2)=4\pi(1-\alpha)/k=4\pi$.

\begin{figure}[t]
\centering
\includegraphics[width=0.3\textwidth]{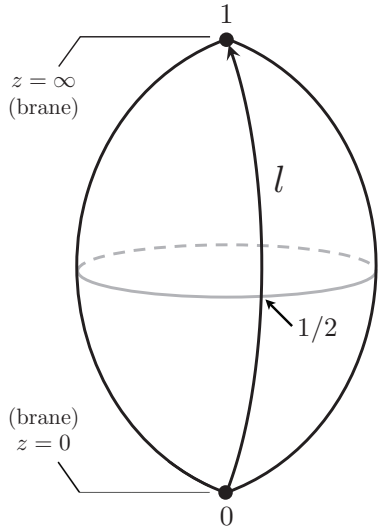}
\caption{``Football"-shaped sphere with two branes at $z=0, \infty$ and the $l$-coordinate.}
\end{figure}

Our goal {in} this subsection is {to derive} zero mode wavefunctions analytically.
Hence, we need to specify the background of gauge fields at first.
Hereafter, we focus on a single abelian gauge field $A$ for simplicity.
If $S^2$ has a smooth metric, $S^2$ is a K\"{a}hler manifold and a solution of the Maxwell equation is realized as a Chern connection of some holomorphic line bundle.
{Thus,} it is natural to assume that the solution $A$ is {given as}
\begin{gather}
	A=i\p\log h\,dz-i\bar\partial \log h\,d\bar{z}
\end{gather}
for some real function $h$, even when the metric has singularities at $z=0$ and $z=\infty$.\,\footnote{
Although we can always add the Wilson line {on the bulk} to the solution $A$ without changing $F$, there {is no constant Wilson line} because $S^2$ is simply connected.
{On another hand,} it is possible to {introduce} the singular Wilson line, such as {$dz/z + d\bar{z}/\bar{z}$}, because $z=0$ and $z=\infty$ are punctures due to {the presence of } branes {effectively}.
However, we do not {take into account} the singular Wilson lines {in this paper and leave such extensions for next projects.}}
Then the {Maxwell equation} $F=f\,{\rm vol}(S^2)$ reduces to the Poisson equation of $h$\,:
\begin{gather}
	-4\p \bar \partial \log h=f G^2.
\end{gather}
This equation is similar to the Liouville equation.
Comparing these two equations, we {obtain}
\begin{gather}
	\log h=\frac{f}{k}\log G+{\rm Re}\,\Omega
\end{gather}
as a general solution of the Poisson equation, where $\Omega$ is some holomorphic function.
{Since} $\Omega$ can be gauged away so that we set $\Omega=0${, the gauge field $A$ under consideration reads}
\begin{gather}\label{gauge_twobranes}
	A=\frac{M}{2}\left(i\p \log G\,dz-i \bar{\partial}\log G\,d\bar{z}\right).
\end{gather}
Here we define $M \equiv 2f /k$.
This solution is consistent with results in supergravity formulation \cite{Lee:2005az,Ludeling:2006hh}.
The spin connection $\omega$ is {expressed as}
\begin{gather}\label{spin_twobranes}
	\omega=i\p \log G\,dz-i \bar{\partial}\log G\,d\bar{z}.
\end{gather}
{Accordingly,} the gauge covariant derivative is $D=d-\omega E-iAQ$ {with plugging Eqs.\,\eqref{gauge_twobranes} and \eqref{spin_twobranes}}, where $E$ is the generator of $SO(2)$ along the extra dimensions and $Q$ is the generator of $U(1)$.
Due to the {Dirac} quantization condition, the product of $U(1)$ charge $q$ and $M$ must be an integer, $qM\in \mb{Z}$.

{First,} let us consider a complex scalar field $\phi$.
The lowest mode equation for $\phi$ is {given by} either of $D \phi=0$ or $\bar{D} \phi=0$.
Their solutions and {the} {lowest bulk KK} masses are obtained as
\begin{gather}\begin{array}{rcll}\label{SCALAR}
	\bar{D}\phi=0 &\Rightarrow& \displaystyle \phi=G^{\frac{qM}{2}}u,&\displaystyle\hspace{0.4cm} m_{\rm KK,\,bulk}^2=\frac{kqM}{2}\vspace{0.3cm}\\
	D \phi=0&\Rightarrow&\displaystyle \phi=G^{-\frac{qM}{2}}\bar{u},&\displaystyle\hspace{0.4cm} m_{\rm KK,\,bulk}^2=-\frac{kqM}{2},
\end{array}\end{gather}
where $u$ is a holomorphic function {of $z$}.
Note that KK masses consist of bulk and brane induced KK masses, which are generated due to branes.
{The} bulk KK masses are common between {the} lowest modes of complex scalar and vector fields respectively.
On another hand, the vector field acquires an additional mass due to the presence of branes, as discussed in the subsection \ref{BraneInducedMass}.
{In normalizing {scalar} wavefunctions,} the norm is given by
\begin{gather}
	\braket{\phi|\phi}=\int G^{qM}|u|^2{\rm vol}(S^2)\hspace{0.5cm}{\rm or}\hspace{0.5cm}\braket{\phi|\phi}=\int G^{-qM}|u|^2{\rm vol}(S^2)
\end{gather}
for each case and must be finite {for normalizable wavefunctions}.
{The normalizability condition $\braket{\phi|\phi}<\infty$ can hold for either of $D \phi=0$ or $\bar{D} \phi=0$} if $qM\neq 0$.

To discuss a spinor wavefunction, we need {to look at} the Dirac operator {on $S^2$ with two branes.}
The Dirac operator $\slashed{D}=\sigma^i D_i$ is {written down as}
\begin{gather}
	\slashed{D}=\frac{2}{G}\begin{pmatrix}& \p -\p\log G^{\frac{-qM-1}{2}} \\ \bar{\partial}- \bar{\partial} \log G^{\frac{qM-1}{2}} & \end{pmatrix}
\end{gather}
and the general solution of the zero mode equation $\slashed{D}\psi=0$ is {obtained as}
\begin{gather}\label{SPINOR}
	\psi=\begin{pmatrix}\psi_- \\ \psi_+ \end{pmatrix}=\begin{pmatrix}G^{\frac{qM-1}{2}}u_- \\[5pt] G^{\frac{-qM-1}{2}}\bar{u}_+\end{pmatrix},
\end{gather}
{where} $u_\mp$ are holomorphic functions.
Its norm $\braket{\psi|\psi}$ is {computed as}
\begin{gather}
	\braket{\psi|\psi}=\int \left(G^{qM-1}|u_-|^2+G^{-qM-1}|u_+|^2\right){\rm vol}(S^2),
\end{gather}
which {has to be} finite {in order to obtain a normalizable wavefunction}.

{Although} abelian gauge fields are not charged under other {$U(1)$ gauge group}, a part of a {non-abelian} gauge field can be charged under {the} Cartan {direction out of YM gauge} group.
The lowest mode {equations are expressed as} $DB_{\bar{z}}=\bar D B_z=0$ and the solutions and {bulk KK masses} are derived {in the same manner as the previous subsection} as
\begin{gather}\begin{array}{rl}\label{VECTOR}
	B_z=G^{\frac{qM}{2}}u_z,&\displaystyle\hspace{0.4cm} m_{\rm KK,\,bulk}^2=-\frac{kqM}{2}, \vspace{0.3cm}\\
	B_{\bar z}=G^{-\frac{qM}{2}}\bar{u}_{\bar{z}},&\displaystyle\hspace{0.4cm} m_{\rm KK,\,bulk}^2=\frac{kqM}{2},
\end{array}\end{gather}
where $u_{z,\,\bar{z}}$ are holomorphic functions.
{Either of $B_z$ and $B_{\bar z}$ must vanish due to the normalizability.}
The norm of $B$ is
\begin{gather}
	\braket{B|B}={2}\int (G^{qM-2}|u_z|^2+G^{-qM-2}|u_{\bar{z}}|^2){\rm vol}(S^2),
\end{gather}
which {has to be} finite for a normalizable mode.
{Norms of scalars {(spinors and vectors)} contain $G^{\pm qM}$ {($G^{\pm qM-1}$ and $G^{\pm qM-2}$)}.}
This difference of powers of $G$ is {due to the difference of spin among the fields}.

In any {case}, the normalizability conditions {seem equivalent up to the power of $G$}.
{Hence,} it is {sufficient} to discuss a simplified condition for a holomorphic function $u$ whose norm is defined by
\begin{gather}
	\braket{u|u}_{\rm hol}\equiv\int G^w|u|^2{\rm vol}(S^2)
\end{gather}
for some integer $w$.
The wavefunction {that we derived so far} is not $u$ itself, but $G^{s}u$ or $G^{s}\bar{u}$ for some (half-)integers $s$.
Substituting the explicit form of $G$ into the above, we obtain
\begin{gather}
	\braket{u|u}_{\rm hol}=\left(\frac{2-2\alpha}{\sqrt k}\right)^{w+2}\int_0^{2\pi} d\theta\int_0^\infty rdr\,\left(\frac{r^{-\alpha}}{1+r^{2-2\alpha}}\right)^{w+2}|u|^2.
\end{gather}
When $u$ has a pole on $\mb{C}{\setminus} \{0\}\subseteq S^2$, this integration diverges so that $u$ can not have poles on $\mb{C} {\setminus} \{0\}$.
If the integration diverges, the divergence {originates} from $r=0$ or $r=\infty$.
{Then,} contributions from $z=0$ and $z=\infty$ {are} evaluated as
\begin{gather}
	\int_0 dr\,r^{-\alpha(w+2)+1}|u|^2\hspace{0.5cm}{\rm and}\hspace{0.5cm}\int^\infty dr\,r^{(w+2)(\alpha-2)+1}|u|^2.
\end{gather}
Obviously, $\braket{u|u}_{\rm hol}$ is finite if and only if these two {terms} are finite.
This condition {strongly} restricts the {possible} power of {$z$ in $u$}, and when we take {an ansatz} $u=z^i$ with an integer $i$, the possible values of $i$ is turned out to be
\begin{gather}\label{2NORMALIZABLE}
	\fl{\alpha\frac{w+2}{2}}\leq i \leq w-\fl{\alpha\frac{w+2}{2}},
\end{gather}
{where} $\fl{t}$ {denotes} an integer part in a real number $t$.
In fact, we can evaluate $\braket{u|u}_{\rm hol}$ for $u=z^i$ analytically {as}
\begin{gather}
	\braket{z^i|z^i}_{\rm hol}=\frac{4\pi(1-\alpha)}{k}\left(\frac{2-2\alpha}{\sqrt k}\right)^w B\left({\frac{2i+2-\alpha(2+w)}{2-2\alpha}}_,\,\,\frac{2w+2-2i-\alpha(w+2)}{2-2\alpha}\right).
\end{gather}
Here $B(t,t')$ is the beta function, and this is finite if and only if $i$ satisfies Eq.\,\eqref{2NORMALIZABLE}.
{Since it} is easy to show $\braket{z^i|z^j}_{\rm hol}=0$ for $i\neq j$,
polynomials of $z$
\begin{gather}\label{NORMALIZABLEHOL}
	z^{i-1+\fl{\alpha\frac{w+2}{2}}}\hspace{0.5cm}\left(i=1,\dots,i_{\rm max}\right)
\end{gather}
provide linearly independent normalizable wavefunctions, where we define $i_{\rm max}$ by
\begin{gather}
\label{DEGENERACY}
	i_{\rm max}=w+1-2\fl{\alpha\frac{w+2}{2}}.
\end{gather}
In Appx.\,B, we show {counting formulae (\ref{COUNTING}) for the number of the linearly independent lowest modes}.
Moreover, the number of modes coincides with $i_{\rm max}$ on two branes background.
Therefore, we conclude that polynomials in Eq.\,\eqref{NORMALIZABLEHOL} give the {orthonormal} linearly independent basis of wavefunctions, and summarize weights $s$ and $w$, the degeneracy $i_{\rm max}$ and the square of the lowest {KK} mass in Tab.\,2.
{Note that {in general, the lowest states of vector fields do not have the same masses due to} brane induced ones.
In other words, for vector fields, $i_{\rm max}$ is not necessarily equal to the number of modes appearing in the low energy effective theory. }
This is because some modes of the vector field get additional mass contributions, depending on their wavefunction property, as discussed in detail in the subsection 3.4.

\begin{table}[t]
\centering
\footnotesize
\begin{tabular}{|cc|C{0.07}|C{0.08}|C{0.21}|c|C{0.21}|c|} \hline
                                            \multicolumn{2}{|c|}{}                                                                                       &    \multirow{2}{*}{$s$}      & \multirow{2}{*}{$w$}&                                   \multicolumn{2}{c|}{$qM>0$}                              &        \multicolumn{2}{c|}{$qM<0$}         \\  \cline{5-8}\rule[-7pt]{0pt}{21pt}
				                                          &                                                                                                              &                                         &&      $i_{\rm max}$       &     $m_{\rm KK,bulk}^2$&          $i_{\rm max}$       &      $m_{\rm KK,bulk}^2$    \\ \hline\hline \rule[-7pt]{0pt}{21pt}
				\multirow{2}{*}{Scalar}        &                                    $\bar{D}\phi=0$                              &        $\frac{qM}{2}$	     &$qM$&         $|qM|+1-2\fl{\alpha\frac{|qM|+2}{2}}$            &    $\frac{k|qM|}{2}$                                          &           ---             &          ---                                                 \\ \rule[-7pt]{0pt}{21pt}
			        	                                  &                                     $D \phi=0$                                       &         $\frac{-qM}{2}$         &$-qM$&         ---                           &   ---                                                                  &           $|qM|+1-2\fl{\alpha\frac{|qM|+2}{2}}$              &           $\frac{k|qM|}{2}$                                           \\ \hline\rule[-7pt]{0pt}{21pt}
				\multirow{2}{*}{Spinor}       &  \multirow{2}{*}{$\begin{pmatrix} \psi_- \\ \psi_+ \end{pmatrix} $} &        $\frac{qM-1}{2}$	     &$qM-1$&         $|qM|-2\fl{\alpha\frac{|qM|+1}{2}}$                &                 $0$                                                  &           ---                            &          ---                                                                    \\ \rule[-7pt]{0pt}{21pt}
			        	                                  &                                                                                                              &        $\frac{-qM-1}{2}$     &$-qM-1$&          ---                          &                           ---                                              &           $|qM|-2\fl{\alpha\frac{|qM|+1}{2}}$                 &          $0$                                                                 \\ \hline\rule[-7pt]{0pt}{21pt}
				\multirow{2}{*}{Vector}       &                                     $B_z$                                                               &        $\frac{qM}{2}$         &$qM-2$&         $|qM|-1-2\fl{\alpha\frac{|qM|}{2}}$            &    $-\frac{k|qM|}{2}$                                         &           ---                            &         ---                                                                 \\ \rule[-7pt]{0pt}{21pt}
			        	                                  &                                     $B_{\bar z}$                                                     &        $\frac{-qM}{2}$         &$-qM-2$&         ---                       &              ---                                                        &           $|qM|-1-2\fl{\alpha\frac{|qM|}{2}}$              &       $-\frac{k|qM|}{2}$\\  \hline
\end{tabular}
\caption{{Possible combinations of} the weights $s$ and $w$, the degeneracy $i_{\rm max}$ and the square of the lowest bulk {KK} mass spectrum {for normalizable wavefunctions}.}
\label{table_twobranes}
\end{table}

\subsection{Three or more branes}
When there {exist} $N$ branes, the Einstein equation reduces to
\begin{gather}
\label{LIOUVILLE}
	-\frac{4}{G^2}\p \bar{\partial} \log G=k+2\pi\sum_{a=1}^N\alpha_a {\delta^2 (z-z_a)}.
\end{gather}
The existence of solutions for the  equation is discussed in Ref.\,\cite{10.1093/imrn/rnv300} for generic $N$.
There {is no solution for $N=1$} and an explicit solution for $N=3$ is {investigated} in Ref.\,\cite{Redi:2004tm}.
Although an explicit solution for $N>3$ {has not been} known, we proceed our discussion {generically}.

It is {easy to} extend our previous {discussion} for the case with $N$ branes.
{Indeed}, when the metric $G$ satisfies Eq.\,\eqref{LIOUVILLE}, a vector potential given as Eq.\,\eqref{gauge_twobranes} {satisfies again} the Maxwell equation {even for the arbitrary number} of branes $N$ and their positions $z_a$.
{It is also quite straightforward to derive the lowest mode wavefunctions analytically, this is because what to do here is to replace the metric $G$ on $S^2$ with two branes with the solution $G$ of Eq.\,\eqref{LIOUVILLE}.}
The general solutions of the lowest mode equations given in Eqs.\,\eqref{SCALAR}, (\ref{SPINOR}) and (\ref{VECTOR}) are {still} valid {as formal expressions}.
Hence all we have to do is to discuss the normalizable condition.

Let {us} consider a holomorphic function $u$ whose norm is defined by
\begin{gather}
	\braket{u|u}_{\rm hol}\equiv\int G^w|u|^2{\rm vol}(S^2)
\end{gather}
for some integer $w$.
When the Liouville equation is satisfied, for each brane on $z_a \,\, (\neq \infty)$, there is a {positive} continuous function $H_a$ such that $H_a(z_a)\neq 0$ and $G$ is expressed as
\begin{gather}
	G(z)=H_a(z)|z-z_a|^{-\alpha_a}
\end{gather}
around $z_a$ (see Appx.\,B {in more details}).
Then, the normalizable condition $\braket{u|u}_{\rm hol}<\infty$ is satisfied if and only if $u$ does not have poles except on branes, and the integration around $z=\infty$ is finite and {also in the vicinity of $z = z_a$,}
\begin{gather}
	\frac{i}{2}\int_{|z-z_a|<\epsilon_a} |z-z_a|^{-\alpha(w+2)}|u|^2\,dzd\bar{z}<\infty
\end{gather}
{is satisfied for small $\epsilon_a>0$.}
When the Laurent expansion of $u$ around $z_a$ starts from $(z-z_a)^{n_a}$, this finiteness is equivalent to
\begin{gather}
	n_a\geq \fl{\alpha_a\frac{w+2}{2}}.
\end{gather}
{Defining} $u_0$ as
\begin{gather}
	u_0(z)=z^{-1}\prod_{a=1}^N(z-z_a)^{\fl{\alpha_a\frac{w+2}{2}}},
\end{gather}
the integration of $\braket{z^i u_0|z^i u_0}_{\rm hol}$ for a positive integer $i$ around $z=\infty$ is finite if and only if
\begin{gather}
	i\leq i_{\rm max}=w+1-\sum_{a=1}^N\fl{\alpha_a\frac{w+2}{2}}.
\end{gather}
Since polynomials $z^i$ are linearly independent, {$z^i u_0$} are also linearly independent.
Moreover, $i_{\rm max}$ is equal to the {number} of normalizable wavefunctions (see Eq.\,\eqref{COUNTING}).
Therefore, we conclude that
\begin{gather}
	z^iu_0\hspace{0.5cm}(i=1,\dots,i_{\rm max})
\end{gather}
spans a basis of normalizable wavefunctions.
When the $N$-th brane is placed {at} $z_N=\infty$, all we have to do is to change $u_0$ by
\begin{gather}
u_0 = z^{-1}\prod_{a=1}^{N-1}(z-z_a)^{\fl{\alpha_a\frac{w+2}{2}}}.
\end{gather}
Although {an} effect {of the $N$-th brane} on wavefunctions seems to {disappear}, the normalizable condition around $z=\infty$ is modified because of the existing brane.
{In the next subsection, we show profiles of the lowest modes which are obtained in this manner.}

\subsection{Wavefunction profiles}
When there are two or three branes, we {analytically} have the solution of $G$ so that we can see explicit profiles of wavefunctions in {those cases}.

First of all, we {show} a case with two branes.
{As slightly shown in the introduction, (bulk) $U(1)$ symmetric} Yukawa couplings {in} the low energy effective theory are proportional to overlap integrals,
\begin{gather}
	\int (\psi^i)^\dagger \psi^j\phi^k\,{\rm vol}(S^2)=\frac{4(1-\alpha)^2}{k}\int_0^\infty rdr\int_0^{2\pi}d\theta\frac{r^{-2\alpha}}{(1+r^{2-2\alpha})^2}(\psi^i)^\dagger \psi^j\phi^k.
\end{gather}
An additional weight function {related to $r$} in the right hand side {originates} from the volume form ${\rm vol}(S^2)$.
To evaluate this integral, {it is useful to utilize} a modified $l$-coordinate\,:
\begin{gather}\label{MODIFIEDL}
	l=1-\frac{1}{1+r^{2-2\alpha}}.
\end{gather}
In fact, its differential $dl$ is {obtained as}
\begin{gather}
	dl=\frac{2(1-\alpha)r^{1-2\alpha}}{(1+r^{2-2\alpha})^2}dr
\end{gather}
and hence {this} absorbs the weight function appearing in the overlap integrals.
{As we will explain later in {the section} \ref{YukawaCoupling}, due to the backgrounds' invariance} under rotations around the axis of $S^2$, non-vanishing {contributions depend} on only $\theta=0$,
\begin{gather}
	\left. \frac{4\pi(1-\alpha)}{k}\int_0^1dl\, \bigl|\psi^i\psi^j\phi^k \bigr| \right|_{\theta = 0}.
\end{gather}
The profiles of absolute values of spinor wavefunctions $|\psi^i|$ {with $U(1)$ charge $qM=27$ and two branes of tensions $\alpha_1 = \alpha_2 = 0.9$} are shown in Fig.\,3 {in terms of the modified $l$-coordinate.}
In Fig.\,3, absolute values for only $\theta=0$ {are} shown because {such} absolute values of the wavefunctions on $S^2$ with two branes do not depend on $\theta$.
{It is interesting to note that} some normalizable wavefunctions on the pure sphere become non-normalizable due to the singularities produced by branes.
This effect is captured by the negative contribution of the degeneracy given in Eq.\,\eqref{DEGENERACY}, which vanishes when we set $\alpha=0$.
{For illustration, we show wavefunction profiles in Fig.\,3 for the brane tension $\alpha=0.9$ and flux $qM=27$}, where there are only three spinor zero modes {correspondingly to three generations.}
This means that branes make 24 zero modes non-normalizable, {compared with the pure sphere setup where there are originally 27 normalizable zero modes due to $qM = 27$}.
Not only the degeneracy, the profiles of holomorphic part of wavefunctions are also affected by branes.
The SM {of particle physics} has three generation matters, so that we focus on {the} flux {patterns} which give $i_{\rm max}=3$.
When there {is no additional brane (equivalent to the pure sphere in the section 2), holomorphic parts of spinor wavefunctions are} $z^{i-1}\,\,(i=1,2,3)$.
On the other hand, when there are two branes {with brane tensions $\alpha_1 = \alpha_2 =0.9$ and flux $qM=27$,} holomorphic parts of spinor wavefunctions are $z^{i+11}\,\,\,(i=1,2,3)$.
{Comparing no brane setup {with} two-brane setup, we can observe that an additional power $z^{12}$ appears in {two-brane} setup.
This additional power makes wavefunction {profiles more localized} on some points, as {shown in} Fig.\,3.
{It is expectable that} overlap integrals $\int(\psi^i)^\dagger\psi^j\phi^k\,{\rm vol}(S^2)$ {can} become {sufficiently} hierarchical {reproducing quark and charged lepton masses, for instance}.}
By comparing Fig.\,3 with Fig.\,1, we can see that the wavefunctions are {peaked at} different {positions} due to the presence of the branes.
{In particular}, for fixed $\phi^k$, overlap integrals {can} become much more hierarchical because of {strong localizations of wavefunctions, and then} an overlap integral {of the mode functions localized around the north pole and around the south pole} becomes much smaller due to the branes.\,\footnote{
We can check that the bosonic wavefunction $|\phi^k|$ is also localized {at} some point due to branes.}

\begin{figure}[t]
\label{profile_twobranes}
\centering
			\includegraphics[width=0.5\textwidth]{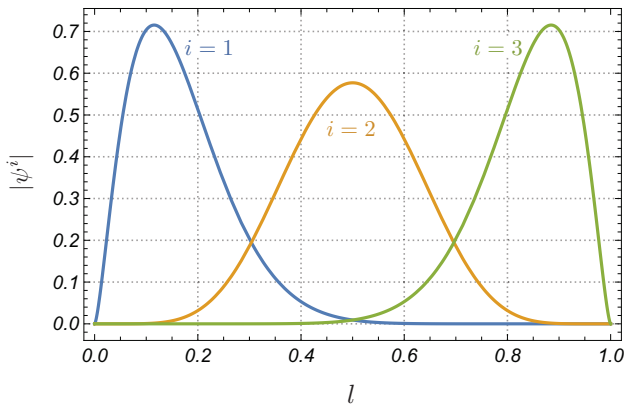}
			\caption{Profiles of absolute values of spinor wavefunctions on the $S^2$ with two branes in {terms of the modified} $l$-coordinate.
			We {set} $\alpha=0.9$ and $qM=27$.
			The constant curvature is given as $k = 1 - \alpha$ so that the volume is normalized as $\int {\rm vol}(S^2)=4\pi$.}
\end{figure}

Next, we would like to discuss a dependence of brane tensions in Fig.\,\ref{branedep}.
In the left and right panels, we {set} $\alpha=0.2$, $qM=3$ and $\alpha = 0.8$, $qM = 11$, respectively.
These setups can be compared with the result of Ref.\,\cite{Conlon:2008qi}, as in Fig.\,1.
It is found that the three peaks of mode functions get closer to $l = 0, 1/2, 1$, as the brane tensions become large.
This is because we can expand the zero mode wavefunctions around $z=0, \infty$, i.e., $l=0, 1$ with the brane tension appeared in powers of $z$, and also the mode peaked around $l=1/2$ tends to get far from the north and south poles with large brane tensions.

Note that, in addition to the rotational symmetry around an axis which penetrates $z=0, \infty$, the two brane background has $\mb{Z}_2$ symmetry generated by $z\rightarrow 1/z$.
Then the lowest mode equation is invariant under this transformation and wavefunctions obey some representations of $\mb{Z}_2$.
In Fig.\,3 and Fig.\,4, the first mode $\psi^1$ and the third mode $\psi^3$ form a doublet and the second mode $\psi^2$ is a singlet.\,\footnote{
	The rotational and the $\mb{Z}_2$ symmetry do not commute each other.
	Now we use a basis of wavefunctions which diagonalize the rotational symmetry, and we can  not diagonalize the $\mb{Z}_2$ symmetry simultaneously.
	Therefore $\psi^1$ and $\psi^3$ form an irreducible $\mb{Z}_2$ doublet even though a pure $\mb{Z}_2$ doublet is reducible.
}
These are the reasons why $|\psi^1|$ and $|\psi^2|$ are mirror images of each other and $|\psi^2|$ is symmetric with respect to $l=1/2$.

\begin{figure}[t]
\centering
\begin{minipage}[t]{\textwidth}
\begin{minipage}[t]{0.5\textwidth}
\includegraphics[width=0.9\textwidth]{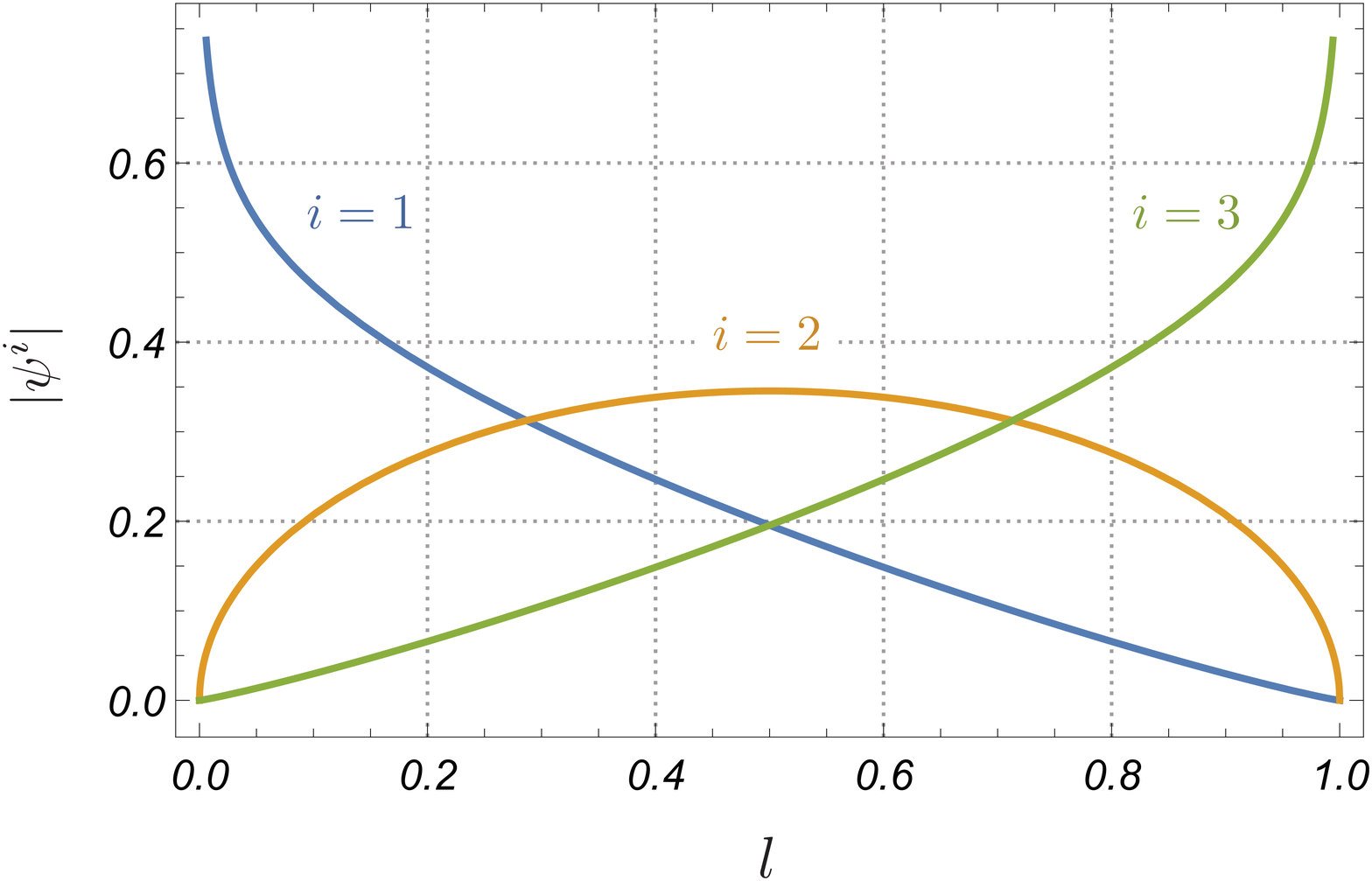}
\end{minipage}
\begin{minipage}[t]{0.5\textwidth}
\includegraphics[width=0.9\textwidth]{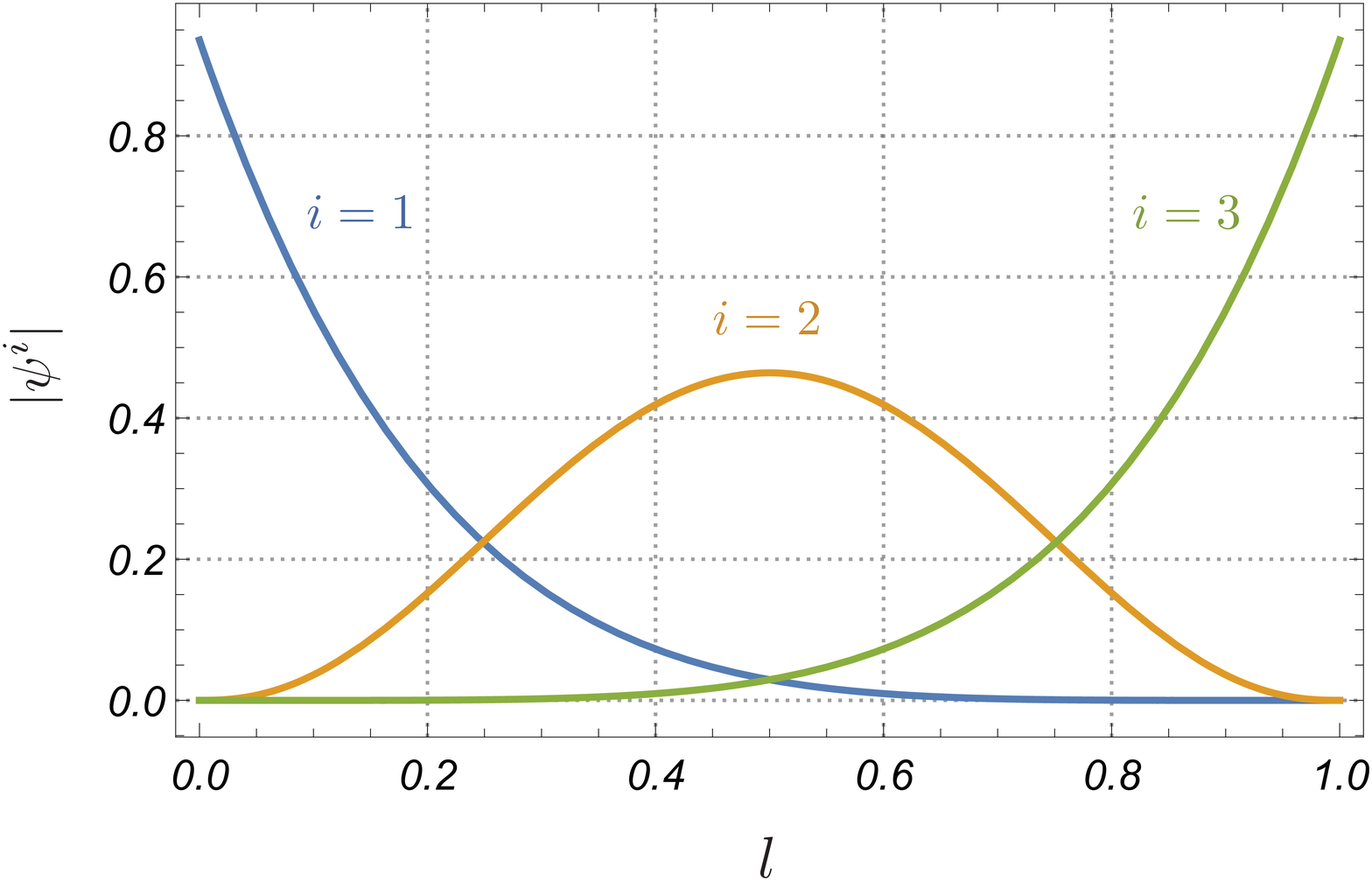}
\end{minipage}
\end{minipage}
\caption{Profiles of absolute values of spinor wavefunctions on the $S^2$ with two branes in {terms of the modified} $l$-coordinate. In the left and right panels, we {set} $\alpha=0.2$, $qM=3$ and $\alpha = 0.8$, $qM = 11$, respectively. It can be seen that different combinations of flux density and brane tension give different localization profiles of the mode functions.}
\label{branedep}
\end{figure}

{In the rest of this subsection, we discuss the wavefunctions on $S^2$ with three branes.}
The solution $G$ of Eq.\,\eqref{LIOUVILLE} for $N=3$ is obtained in Ref.\,\cite{Redi:2004tm} as
\begin{gather}
	G=\frac{2}{\sqrt k}\frac{|\Phi'|}{1+|\Phi|^2},
\end{gather}
where $\Phi$ is a multivalued holomorphic function
\begin{gather}
	\Phi(z)=C \, \frac{{}_2F_1[a_1,b_1,c_1:z]}{{}_2F_1[a_2,b_2,c_2:z]}z^{1-\alpha_1}.
\end{gather}
Although $\Phi$ itself is multivalued, {it is easy to confirm that} $G$ is single-valued.
${}_2F_1[a,b,c:z]$ is the hypergeometric function normalized as ${}_2F_1[a,b,c:0]=1$.
Real constants $a_i,\,b_i$ and $c_i$ {in the hypergeometric function} are given by
\begin{gather}\begin{array}{rclrcl}
	a_1&=&\displaystyle \frac{2-\alpha_1+\alpha_2-\alpha_3}{2}, & \hspace{0.3cm}a_2&=&\displaystyle \frac{\alpha_1+\alpha_2-\alpha_3}{2}, \vspace{0.3cm}\\
	b_1&=&\displaystyle \frac{-\alpha_1+\alpha_2+\alpha_3}{2},&\hspace{0.3cm}b_2&=&\displaystyle \frac{-2+\alpha_1+\alpha_2+\alpha_3}{2}, \vspace{0.3cm}\\
	c_1&=&2-\alpha_1,&\hspace{0.3cm}c_2&=&\alpha_1,
\end{array}\end{gather}
where $\alpha_1$ {($\alpha_2$ and $\alpha_3$) denotes} a brane tension of the brane at $z=0$ {($z = 1$ and $z=\infty$)}.
As shown in Ref.\,\cite{Troyanov:1991}, these three brane tensions give a constant curvature on $S^2$ with three branes as
\begin{gather}
k = \frac12 (2 - \alpha_1 -\alpha_2 -\alpha_3),
\end{gather}
when the 2d volume is fixed as $\int {\rm vol}(S^2) = 4\pi$.
{Here we fix positions of {three} branes on $z=0,\,1$ and $\infty$ by using $SL(2,\mb{C})$ symmetry of $S^2$.}\,\footnote{{In general, arbitrary distinct three points of $S^2$ can be {moved} to other arbitrary distinct three points by an $SL(2,\mb{C})$ transformation, called an M\"{o}bius transformation.
The positions of three branes are uniquely fixed by this transformation.
If we try to introduce the forth brane, we can not fix its position, and hence the position of the forth brane becomes a model parameter.}}
The overall constant $C$ in $\Phi$ is {given as}
\begin{gather}
	C=\frac{\Gamma(a_1)\Gamma(b_1)\Gamma(c_2)}{\Gamma(a_2)\Gamma(b_2)\Gamma(c_1)}\sqrt{-\frac{\cos\pi(\alpha_1-\alpha_2)-\cos\pi\alpha_3}{\cos\pi(\alpha_1+\alpha_2)-\cos\pi\alpha_3}}.
\end{gather}
Note that $C$ becomes imaginary number for some {values of} $\alpha_{1,\,2,\,3}$.
In such a case, the solution $G$ is invalid as pointed out in Ref.\,\cite{Redi:2004tm}.

Similarly to the setup of two branes, we find (quasi-)localization profiles of zero mode wavefunctions in $l$-coordinate direction.
It is also confirmed that stronger brane tensions give more localizing shapes of zero modes on $S^2$ with three branes.
Interestingly, in contrast to $N=2$, it turns out that the wavefunctions possess a phase-dependence in a polar coordinate $z = r e^{i\theta}$.
In Fig.\,\ref{threebranes_theta} where we set $\alpha_1 = \alpha_3 = 0.4$, $\alpha_2 = 0.1$ and $qM=4$, the norms of wavefunctions shows their $\theta$-dependence at $l = 0.9$.
In principle, this $\theta$-dependence can be understood as a breakdown of an isometry on $S^2$, due to the presence of three branes.
It is no longer possible to evaluate the Yukawa couplings only at $\theta = 0$, and it is expectable that the $\theta$-dependence would lead to the CP violation in the low energy effective Yukawa couplings.
We will discuss a relation between the isometry of $S^2$ with branes and the CP violation in Appx.\,C from a mathematical point of view.

\begin{figure}[t]
\centering
	\includegraphics[width=0.5\textwidth]{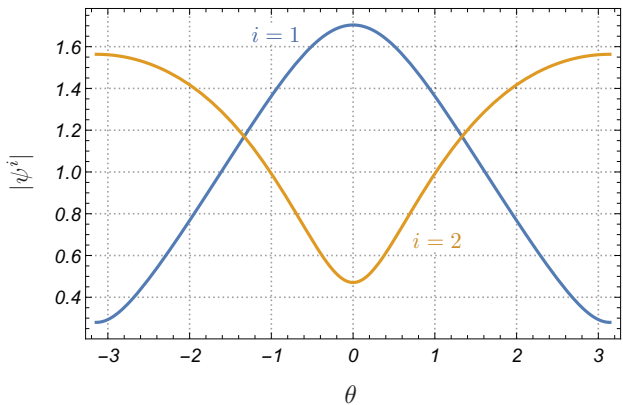}
			\caption{Phase-dependences of the wavefunctions $|\psi^i|$ at $l = 0.9$ on $S^2$ with three branes. Since we set $\alpha_1= \alpha_3 = 0.4$, $\alpha_2=0.1$ and $ qM = 4$, there appear two mode functions $(i = 1, 2)$ where non-trivial $\theta$-dependences are observed and those can give complex-valued entries in Yukawa couplings.
}
\label{threebranes_theta}
\end{figure}

\subsection{Brane induced mass}
\label{BraneInducedMass}
Before closing this section, we need to discuss the KK masses of the lowest states in complex scalar and vector fields.
While spinor fields have massless modes as the lowest states, i.e., the zero modes in any case, the complex scalars and vectors are likely to acquire non-vanishing contributions from the bulk flux (as in the magnetized torus/orbifold \cite{Cremades:2004wa}).
In particular, the lowest mode wavefunctions of vector fields receive brane induced masses coming from Yang--Mills kinetic terms.

Let $\phi$ be a wavefunction of a complex scalar field, normalized as $\int_{S^2} \bar{\phi}\phi\,{\rm vol}(S^2)=1$.\,\footnote{
	Our discussion in this subsection can be easily extended for a general 2d compact manifold if the manifold is not a torus and its metric satisfies the Liouville equation.
}
The KK mass $m_{\rm KK}^2$ of the mode is given by
\begin{gather}
	m_{\rm KK}^2 = \int_{S^2} d_A\bar{\phi}\wedge \star \,d_A\phi \equiv \lim_{\epsilon\rightarrow 0}\,\int_{S^2\backslash \cup_a D_{a,\epsilon}}d_A\bar{\phi}\wedge \star \,d_A\phi,
\end{gather}
where $d_A=d-iqA$ is the covariant exterior derivative on flux background and $\star$ is the Hodge dual on $S^2$.
$D_{a,\epsilon}$ denotes an open disc around a singularity $z_a$ with a small radius $\epsilon>0$.
Without any assumption, we can rewrite the integrand as
\begin{gather}\label{INTEGRAND}
	d_A\bar\phi \wedge \star d_A\phi = \frac{1}{2} d\star d(\bar{\phi}\phi) + qF\bar\phi\phi,
\end{gather}
where we use $\bar{\p}_A\phi=0$.
Here, $F=dA = \tfrac12 kM {\rm vol}(S^2)$ is a field strength associated with flux background.
The second term in Eq.\,\eqref{INTEGRAND} clearly gives bulk mass $m^2_{\rm KK, \, bulk}= k qM/2$.
In short, the total KK mass of the mode is given as
\begin{gather}
	m_{\rm KK}^2 = m^2_{\rm KK, \, bulk}-\lim_{\epsilon\rightarrow 0}\frac{1}{2}\sum_a\oint_{\p D_{a,\epsilon}}\star d(\bar{\phi}\phi).
\end{gather}
Note that the second term may correspond to a contribution of singular curvature around branes and possibly give an infinite value which we regularize or neglect in usual schemes.

Next, we discuss the KK mass in vector fields.
When a vector field is a part of a Yang--Mills field and also some Cartan directions develop their VEVs, the Yang--Mills kinetic term contains
\begin{gather}
\label{VectorKKmass}
	iq\int_{S^2}F\wedge \star(\bar{B}\wedge B).
\end{gather}
This term becomes the KK mass $m^2_{\rm KK}$.
Recalling that the underlying gauge potential is given by $A=-\tfrac12 M \star d\log G$ and also the metric $G$ satisfies the Liouville equation \eqref{LIOUVILLE}, it can be observed that $F=dA$ includes localized (geometric) flux on brane positions,
\begin{gather}
\label{LocalizedFlux}
	F=dA=\left(\frac{kM}{2}+\pi M\sum_{a=1}^N\alpha_a\delta^2(z-z_a)\right){\rm vol}(S^2).
\end{gather}
Substituting Eq.\,\eqref{LocalizedFlux} into Eq.\,\eqref{VectorKKmass}, we obtain
\begin{gather}
	m_{\rm KK}^2=m_{\rm KK, \, bulk}^2+i\pi qM\int_{S^2}\left(\sum_{a=1}^N\alpha_a\delta^2(z-z_a)\right)\bar{B}\wedge B.
\end{gather}
When for $qM>0$, the $\bar{z}$ component $B_{\bar{z}}$ of $B=B_zdz+B_{\bar z}d\bar{z}$ vanishes and $B$ is normalized as $\int\bar{B}\wedge\star B=1$, the bulk mass reads
\begin{gather}
	m_{\rm KK, \, bulk}^2 =-\frac{k qM}{2},
\end{gather}
as shown in the previous subsections.
Recalling the expressions of vector wavefunctions, let $B_z=G^{\frac{qM}{2}}(z-z_a)^{n_a}v_{z,a}$ be a local expression of $B_z$ around $z=z_a$.
Here $v_{z,a}$ is a holomorphic function with taking a finite value at $z_a$, i.e., $v_{z,a}(z_a)\neq 0$ and $n_a$ is the order of zero of the holomorphic part of the wavefunction $B_z$.
Then, the brane induced mass is computed as
\begin{gather}\begin{array}{rl}
	m_{\rm KK, \, brane}^2& \equiv \displaystyle i\pi qM\int_{S^2}\left(\sum_{a=1}^N\alpha_a\delta^2(z-z_a)\right)\bar{B}\wedge B\vs\\
	&=\displaystyle-2\pi qM\sum_{a=1}^N\alpha_a H_a(z_a)^{qM-2}|v_{z,a}(z_a)|^2\lim_{z\rightarrow z_a}|z-z_a|^{2n_a-(qM-2)\alpha_a},
\end{array}\end{gather}
where $G=H_a|z-z_a|^{-\alpha_a}$ is a local expression of the metric around $z_a$.
If some $n_a$ satisfies $n_a< (qM-2)\alpha_a/2$, the brane induced mass gives a negative infinity and should be regularized in appropriate manners.
For $n_a = (qM-2) \alpha_a/2$, the brane induced mass takes a finite non-vanishing values.
By repeating the above discussion also for non-vanishing $B_{\bar z}$ and $qM<0$, the brane induced mass generically reads
\begin{gather}
m_{\rm KK,\,brane}^2 = -2\pi|qM|\sum_{a=1}^N\alpha_aH_a(z_a)^{|qM|-2}|v_{z,a}(z_a)|^2\delta_{2n_a-(|qM|-2)\alpha_a,0},
\end{gather}
and this gives non-zero mass only for $n_a = (|qM|-2)\alpha_a/2$.

\section{Yukawa couplings on $S^2$ with flux and branes}
\label{YukawaCoupling}
{In this section, we comment on the effective Yukawa couplings by overlap integrals and their property related to a non-vanishing CP phase.}

{Let us consider} $S^2$ with two branes whose dimensionless tensions are {given by $\alpha \,\, (=\alpha_1 = \alpha_2)$}.
We derived wavefunctions analytically in {the subsection} \ref{TWOBRANES} and overlap integrals are {expressed as}
\begin{gather}\label{OVERLAP2}
	\int (\psi^i)^\dagger\psi^j\phi^k\,{\rm vol}(S^2)=\frac{i}{2}\mc{N}\int G^{w+2}\bar{z}^nz^m  \,dzd\bar{z},
\end{gather}
for some weight $w$, which is determined by $U(1)$ charges and spins of {matter fields}, and {for} integers $n$ and $m$, which are determined by the mode labels $i,\,j$ and $k$ {among} wavefunctions.
$\mc{N} \,\, (>0)$ is a product of normalization factors for {each wavefunction}.
Since $G$ is independent on an angular coordinate $\theta$, {one can find}
\begin{gather}\label{ORTHOGONAL}
	\int (\psi^i)^\dagger\psi^j\phi^k\,{\rm vol}(S^2)=\mc{N}\left(\int_0^\infty dr\, G^{w+2}r^{n+m+1}\right)\left(\int_0^{2\pi}d\theta \,e^{i(m-n)\theta}\right)
\end{gather}
vanishes when $m\neq n$.
By inserting the explicit form (\ref{TWOBRANESBACKGROUND}) of $G$ into the first factor and introducing {the} modified $l$-coordinate (\ref{MODIFIEDL}), we obtain
\begin{gather}
	\int_0^\infty dr\, G^{w+2}r^{2n+1}=\frac{2-2\alpha}{k}\left(\frac{2-2\alpha}{\sqrt k}\right)^{w}B\left({\frac{2n+2-\alpha(w+2)}{2-2\alpha}}_,\,\,\frac{2w+2-2n-\alpha(w+2)}{2-2\alpha}\right),
\end{gather}
where $B(t,\,t')$ is the beta function.
{The} products of normalization factors $\mc{N}$ is also written by the beta function, so that we can {analytically} write down the Yukawa couplings.

Although {the analytic forms of the Yukawa couplings are realizable in principle}, the result {may be} phenomenologically unfavorable.
Recall that holomorphic parts of wavefunctions in Eq.\,\eqref{OVERLAP2} are {written as}
\begin{gather}
	\psi^i\propto z^{i-1+\fl{\alpha\frac{(qM)_1+1}{2}}},\hspace{0.3cm} \psi^j\propto z^{j-1+\fl{\alpha\frac{(qM)_2+1}{2}}},\hspace{0.3cm}\phi^k\propto z^{k-1+\fl{\alpha\frac{(qM)_3}{2}}},
\end{gather}
where $(qM)_{1,\,2,\,3}$ are the numbers of flux which {three matters} feel.
These {expressions} and $\int e^{i(m-n)\theta}\,d\theta$ in Eq.\,\eqref{ORTHOGONAL} {imply that} non-vanishing Yukawa couplings {for} fixed $\phi^k$ appear only when
\begin{gather}
	i-j={({\rm const})}.
\end{gather}
The full rank Yukawa matrix is phenomenologically desired to reproduce massive matters, however, the above Yukawa matrix to $\phi^k$ {can be} full rank only when the right hand side vanishes.
{Then,} a possible full rank Yukawa matrix is a diagonal one at most, and hence there {is} no mixing between matters and {it is not suitable for reproducing the SM quark mixing angles, for instance}.
In addition, overlap integrals given in Eq.\,\eqref{OVERLAP2} {take only real values}, so that the Yukawa matrix does not {break} CP symmetry.
{This is again not suitable for realizing realistic quark flavor structures.}

The origin of these phenomenologically unsatisfactory points is a symmetry of the background.
In fact, the background metric $G^2dzd\bar{z}$ and the magnetic flux $F$ {are} invariant under rotations around an axis which penetrates two points $z = 0, \infty$.\,\footnote{
	Although the background metric has isometry, it is not necessarily symmetry of magnetized models.
	For example, translations on the flat torus are isometries, however, any translations can not keep magnetic fluxes invariant and magnetic fluxes break the symmetry of the torus.
	On the other hand, isometry of $S^2$ becomes symmetry of the extra dimensional model even when magnetic fluxes are turned on (see Appx.\,C).
}
Therefore a finite dimensional vector space $V$ of wavefunctions, for {example}, spinor wavefunctions which feel the flux $qM$, is closed under the isometry group $U(1)_{\rm isom}$ and $V$ becomes a unitary representation of $U(1)_{\rm isom}$.
Then we have an orthogonal decomposition
\begin{gather}
	V= \bigoplus_{i\in\mb{Z}} V_i
\end{gather}
in terms of $U(1)_{\rm isom}$ charges $i$.
If two wavefunctions carry different $U(1)_{\rm isom}$ charges, they are orthogonal.
Since different modes have different powers of $z$, the dimension of $V_i$ is at most one and the only possible full rank Yukawa matrix is diagonal.
We can also understand the reality of the matrix as a consequence of $U(1)_{\rm isom}$, because the total $U(1)_{\rm isom}$ charge of the integrand {in} Eq.\,\eqref{OVERLAP2} must vanish, and hence the integrand can not depend on the angular coordinate.

Thus it is phenomenologically important to break the isometry of the pure $S^2$, and this is {performed} by introducing additional branes.
{In other words, $S^2$ with three branes described in the subsection 3.3 is promising in concrete model constructions.}
Unfortunately, it is difficult to compute {such} Yukawa couplings analytically {once} three or more branes are introduced.
{Nevertheless}, we can observe {that} the $U(1)_{\rm isom}$ isometry of $S^2$ with two branes are broken down {partly} by the third brane because wavefunctions acquire the angular dependence.
Some brane configuration breaks the isometry $SO(3)$ of the pure sphere into a finite subgroup of the $SO(3)$.
Although magnetic flux is turned on, the finite subgroup becomes symmetry of the model (see Appx.\,C).
Therefore, such a brane configuration can be thought as an origin of {(semi-)realistic flavor structures in} the low energy effective theory.
{For instance, some finite subgroups of $SO(3)$ are isomorphic to symmetric group $S_4$ and alternative groups $A_4$, $A_5$, and these groups are extensively studied as flavor symmetries.}
{We will revisit this topic in future projects.}

\section{Conclusion and discussion}
{In this paper, we have formulated the 6d $U(1)$ gauge theory whose 2d extra dimensions are compactified on the sphere with flux background and multiple branes, and derived the wavefunctions of the zero modes of scalar, spinor and vector fields after KK decomposition.
In counting the number of independent KK zero modes, we have focused on their normalizability and given mathematical formulae based on the index theorem (as mentioned in appendices).
Then, it has been found that such wavefunctions on the sphere with flux and branes possess stronger localization profiles than those on the pure sphere, and also that the profiles on the sphere with three branes are dependent on $\theta$ in expressing a polar coordinate $z = r e^{i \theta}$.
From the viewpoint of overlap integrals in dimensional reduction, the Yukawa couplings can potentially break CP symmetry due to the presence of flux background and three branes.

Before closing this paper, we would like to comment on several applications of our wavefunctions.
Although in this paper, we have not constructed phenomenological models, it is fair to say that our setup, namely the sphere with branes, can have adequate possibilities to construct phenomenologically promising models.
In general, the singularity such as orbifold fixpoints plays an important role on which we put ingredient terms, for instance, the (localized) Yukawa couplings, the localized mass terms and Higgs potential, as investigated in toroidal/orbifold flux compactifications \cite{Buchmuller:2017vho, Buchmuller:2017vut, Ishida:2017avx, Ishida:2018bbl, Abe:2018qbp}.

Another possibility is to introduce localized fluxes at singularities (similarly to the second term in Eq.\,\eqref{LocalizedFlux}, but they should have quantized coefficients).
In the same manner as Refs.\, \cite{Buchmuller:2015eya,Buchmuller:2018lkz,Lim:2018lgg}, it is possible in principle to formulate localized fluxes in terms of the Green function.
It gives non-trivial Wilson-loop around the singularity, which can contribute to total flux density and an appropriate way of the Dirac charge quantization in full setup.
In particular, it is interesting as an application to utilize such generic backgrounds of localized fluxes for Gauge-Higgs unification scenario in which extra dimensional components of vector potential break the electroweak symmetry \cite{Hosotani:1983xw}.
As discussed in Ref.\,\cite{Hatanaka:1998yp}, Wilson-loops control the one-loop induced effective potential and its concrete form.\,\footnote{See Refs.\,\cite{Buchmuller:2016gib,Ghilencea:2017jmh,Buchmuller:2018eog} for magnetized torus case.}
It can be expected that the non-trivial Wilson-loop around the brane singularity would give some correction to effective potential at loop level.
We will leave it for one of the future projects.
}

\section*{Acknowledgment}
The authors would like to thank Keigo Sumita for valuable discussions at the early stage of this project.
S.I. would like to thank Hiroyuki Abe and Yutaka Sakamura for useful comments.
Y.T. would like to thank Wilfried Buchm\"uller and Jakob Moritz for instructive comments on this manuscript.
Y.T. is supported in part by Grants-in-Aid for JSPS Overseas Research Fellow (No.~18J60383) from the Ministry of Education, Culture, Sports, Science and Technology in Japan.

\appendix
\renewcommand{\thesection}{\Alph{section}}
\section{Patches and flux quantization on $S^2$ with flux}
\begin{figure}[t]
\centering
\includegraphics[width=0.85\textwidth]{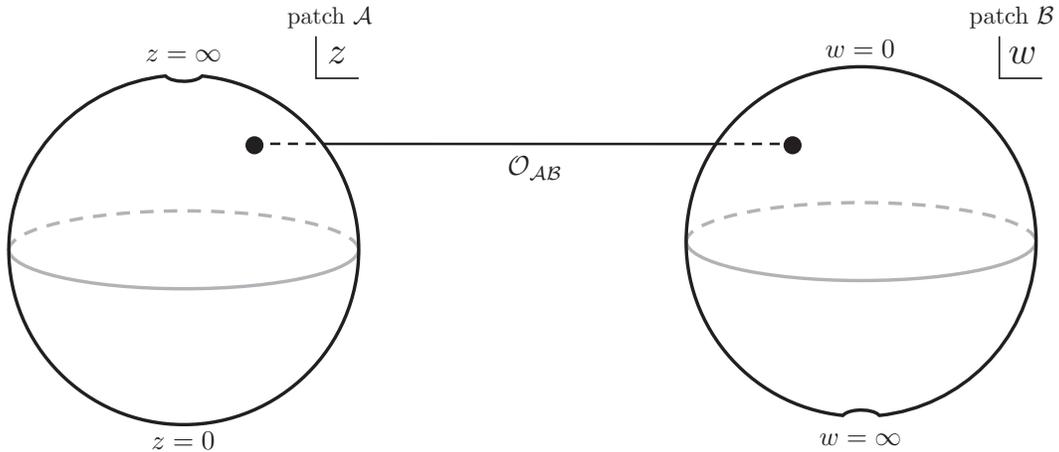}
\caption{Image of two patches covering north and south hemispheres $\cal A, B$ and necessary gauge transformation ${\cal O}_{\cal AB}$.
Coordinates $z$ on $\cal A$ and $w$ on $\cal B$ are connected by $w=-1/z$ on $\cal A\cap B$.}
\label{patch}
\end{figure}

We summarize patches and the {Dirac} quantization of magnetic flux on a {pure} $S^2$ discussed in Ref.\,\cite{Conlon:2008qi}.
$S^2=\mb{C}\cup\{\infty\}$ is covered by two complex planes $\mb{C}$ {(which are {diffeomorphic} to two 2d real planes)}.
The first one, patch $\mc{A}$, covers the lower part of $S^2$ and {another one}, patch $\mc{B}$, does the upper one of $S^2$.
Coordinates on these {two} patches are {connected} by $w=-1/z$ on a common region $\mc{A}\cap \mc{B}$ {(see Fig.\,\ref{patch})}, where $z$ is {a} coordinate on $\mc{A}$ and $w$ is one on $\mc{B}$.
{Unlike Refs.\,\cite{Wu:1976ge, Wu:1976qk} and usual manners, note that the authors in Ref.\,\cite{Conlon:2008qi} take the notation that patches $\cal A$ ($\cal B$) describe $z \in \mathbb{C}\setminus\{\infty\}$ ($z \in \mathbb{C} \setminus \{0\}$, i.e., $w \in \mathbb{C} \setminus \{\infty\}$), respectively.}
At first, we solve the Maxwell equation on the patch $\mc{A}$.\,\footnote{
	We {would like to point out} that a collection of real 1-forms $\{A_i:U_i\rightarrow T^*_\mb{R}S^2\}_i$, where $\{U_i\}$ is an open covering of $S^2$ and $T^*_\mb{R}S^2$ is the (real) cotangent bundle, is a gauge potential of a globally defined real 2-form $F$,  called a field strength, if $dA_i=F$ holds for each $i$.
}
The solution $A_\mc{A}$ is {given as}
\begin{gather}
	A_\mc{A}=-\frac{iM\bar{z}}{2(1+|z|^2)}dz+\frac{iMz}{2(1+|z|^2)}dz.
\end{gather}
This solution is defined on the patch $\mc{A}$ and {ill-defined only} on the north pole {($z = \infty$)}.
To see how to extend $A_\mc{A}$ on entire $S^2$, we represent $A_\mc{A}$ in terms of the coordinate $w$ of $\mc{B}$\,:
\begin{gather}
	A_\mc{A}=\frac{iM}{2(1+|w|^2)}\frac{dw}{w}-\frac{iM}{2(1+|w|^2)}\frac{d\bar{w}}{\bar{w}} \qquad (w\neq0).
\end{gather}
{As one can see directly from this gauge potential, it} tells that {$A_\mc{A}$ is not defined on $w=0$ due to divergence.}
{Then}, we {need to} prepare another gauge field $A_\mc{B}$ defined on the patch $\mc{B}$ {in order for the solution to be globally well-defined.}
Since field strengths of $A_\mc{A}$ and $A_\mc{B}$ must {match consistently}, their difference must be {connected} by a gauge transformation $\mc{O}_{\mc{AB}}$.
We define $A_\mc{B}$ by
\begin{gather}
	A_\mc{B}=A_\mc{A}-id\log\mc{O}_{\mc{AB}}=-\frac{iM\bar{w}}{2(1+|w|^2)}dw+\frac{iMw}{2(1+|w|^2)}d\bar{w},
\end{gather}
where $\mc{O}_{\mc{AB}}$ is {a transition function\,:}
\begin{gather}\label{GAUGETRAFO}
	\mc{O}_{\mc{AB}}=\left(\frac{w}{\bar{w}}\right)^{M/2} \qquad (w\neq0).
\end{gather}
{Although} $A_\mc{A}$ diverges at $w=0$, $A_\mc{B}$ does not and {thus} we can {introduce} $A_\mc{B}$ on {the} entire $\mc{B}$.
If a matter field is charged with {$U(1)$} charge $q$, it receives a gauge transformation $(\mc{O}_{\mc{AB}})^q$ when we move {from the patch $\mc{A}$} to the patch $\mc{B}$.
Due to single-valuedness of the gauge transformation \eqref{GAUGETRAFO}, $qM$ must be an integer under the Dirac quantization \cite{Conlon:2008qi}.

\section{Counting normalizable zero modes}
\label{CountNormalizableMode}

The number of {independent} zero modes {in} 6d models is counted by the Riemann--Roch theorem.
In this appendix, we extend {counting formulae} in the presence of branes.
We do not assume that our extra dimensions are compactified on $S^2$, but on a more general 2d compact surface $X$, which is not a torus.
{The line element {$ds^2$} that we want to discuss} is no longer smooth on brane positions, however, {the bulk part away from the brane positions can be smooth.}
It is assumed that we have {a} complex coordinate $z$ on $X$, {at} each point including a brane position, of $X$ such that {the} metric {$ds^2$} is expressed as
\begin{gather}\label{METRIC}
	ds^2 = \frac{1}{2}G(z)^2(dz\otimes d\bar{z}+d\bar{z}\otimes dz).
\end{gather}

First of all, we mention a singular behavior of {the} metric around a brane.
We take a {sufficiently} small open set $U$ around the brane which does not contain any other brane.
Let $z$ be a complex coordinate on $U$ such that the metric is expressed as in Eq.\,\eqref{METRIC}.
When we place {a} brane at $z=0$, the Liouville equation {reads}
\begin{gather}
	-\frac{4}{G^2}\p\bar\partial \log G=k+2\pi \alpha \delta^2(z),
\end{gather}
where $\alpha$ is a dimensionless brane tension, $k$ is non-vanishing constant curvature of $X$ and $\delta^2(z)$ is a delta function normalized as $\int_U {\delta^2(z)} \varphi(z)\,{\rm vol}(X)=\varphi(0)$ for {any} smooth function $\varphi$ on $U$ whose support is compact.
Note that {the above} equation is not an ordinary differential equation but {should be considered as} an integral equation because the equation contains the delta function.
{Hence,} a precise meaning of the Liouville equation on $U$ should be
{\begin{gather}
	\int_{U\backslash\{0\}}d\varphi\wedge \star d\log G=k\int_{U\backslash\{0\}}\varphi \,{\rm vol}(X)+2\pi \alpha \varphi(0),
\end{gather}
where $d$ is an exterior derivative and $\star$ is the Hodge dual.}
By a straightforward calculation, it reduces to the non-singular Liouville equation
\begin{gather}
-\frac{4}{G^2}\p \bar\partial\log G=k \qquad (z \in U \setminus\{0\})
\end{gather}
and
\begin{gather}
\lim_{\epsilon{\to 0}}\int_0^{2\pi} d\theta\left(-\varphi \,r\p_r\log G\right) \Bigr|_{r=\epsilon} =2\pi\alpha \varphi(0),
\end{gather}
where we use $z=re^{i\theta}$ {and its derivative $\partial_r \equiv \partial/\partial r$.}
When we define a function $H$ on $U\setminus \{0\}$ as $G=Hr^{-\alpha}$, the above equation leads {to}
\begin{gather}\begin{array}{rcll}\label{IDENTITIES}
	-4\p\bar\partial \log H&=&\displaystyle k G^2 &({\rm on}\,\,\,U\setminus\{0\}),\vspace{0.4cm} \\
	\displaystyle\int_0^{2\pi}d\theta\,r\p_r\log H&\rightarrow & 0 &(r\rightarrow 0).
\end{array}\end{gather}
If the volume of $X$ diverges, the divergence comes from brane positions because $X$ is compact and the metric is smooth except on brane positions.
Let $R$ be a positive real number such that $\acom{z\in \mb{C}\mid |z|<R}$ is contained in $U$.
Then, a contribution of the metric around $z=0$ to the volume $\int_X{\rm vol}(X)$ is evaluated by
\begin{gather}
\lim_{\epsilon{\to 0}}\int_\epsilon^R rdr\int_0^{2\pi}d\theta\,G^2,
\end{gather}
{and also turned out to be} finite.
In fact, by using Eq.\,\eqref{IDENTITIES}, we have
\begin{gather}\begin{array}{rcl}
	\displaystyle\lim_{\epsilon{\to 0}}\int_\epsilon^R rdr\int_0^{2\pi}d\theta\,G^2&=&\displaystyle-\frac{1}{k}\lim_{\epsilon{\to 0}}\int_\epsilon^R dr\int_0^{2\pi} {d\theta} \left(\p_r(r\p_r)+\frac{1}{r}\p_\theta^2\right)\log H  \vspace{0.3cm}\\
	&=&\displaystyle -\frac{1}{k}\lim_{\epsilon\to 0}\left[\int_0^{2\pi} {d\theta}\, r\p_r\,\log H\right]_{r=\epsilon}^{r=R}\vspace{0.3cm}\\
	&=&\displaystyle -\frac{1}{k}\int_0^{2\pi} {d\theta}\, r\p_r\log H \Bigr|_{r=R} \\
	&<& \infty.
\end{array}\end{gather}
{Thus}, when {the} metric on {the} compact surface $X$ satisfies the Liouville equation with $k\neq 0$, the volume of $X$ is finite, i.e., $\int_X{\rm vol}(X)<\infty$.
In addition, according to Ref.\,\cite{Chou:1994}, the above finiteness implies that there is a holomorphic function $\Phi$ around $z=0$ satisfying
\begin{gather}
	G=Hr^{-\alpha}=\frac{2}{\sqrt k}\frac{|z\p \Phi+(1-\alpha)\Phi|}{1\pm|\Phi r^{1-\alpha}|^2}r^{-\alpha} \qquad (\Phi(0)\neq 0).
\end{gather}
Here the sign $\pm$ is a sign of $k$.
{In particular, since} $H$ is {a} continuous function defined around $z=0$ and $H(0)\neq 0$, for a {sufficiently} small real positive number $R$, $H$ is bounded by real positive constants from below and above on $\acom{z\in \mb{C}\,|\,\,|z|<R}\subseteq U$.

Let $L\rightarrow X$ be a holomorphic line bundle and $h$ a hermitian metric on $L$.
Although there are branes on $X$, we assume $L\rightarrow X$ is globally defined, and then impose the Maxwell equation on the Chern connection induced by $h$.
Since we have the Liouville equation, the Maxwell equation is solved as
\begin{gather}
	h=\frac{2w}{\chi(X)}\log G,
\end{gather}
where $w\in \mb{Z}$ is the Chern number of $L\rightarrow X$ and $\chi(X)\neq 0$ is the Euler characteristic of $X$.\,\footnote{
	We give a rough sketch of the proof.
	Let $V$ be a sufficiently small open set on $X$ which contains all branes.
	At first, we prepare a smooth K\"{a}hler metric, which is explicitly constructed with a partition of unity such that $X$ coincides with our singular metric on $X \setminus V$.
	The smooth metric makes $X$ be a K\"{a}hler manifold and the Hodge decomposition is available to prove the existence of a hermitian metric $\tilde h$ on $L\rightarrow X$ whose Chern connection solves the Maxwell equation with respect to the smooth metric.
	The local expression of $\tilde h$ on $X \setminus V$ is
	\begin{gather}
		\tilde h = \frac{2w}{\chi(X)}\log G+{\rm Re}\,\tilde \Omega,
	\end{gather}
	because the smooth metric coincides with the original singular metric $G^2 dzd\bar{z}$.
	A holomorphic function $\tilde \Omega$, which makes $\tilde h$ compatible with a transition function of $L\rightarrow X$, can be gauged away by an automorphism of the holomorphic line bundle $L\rightarrow X$.
	Therefore we have $\tilde h=\frac{2w}{\chi(X)}\log G$.
	$\tilde h$ does not satisfy the Maxwell equation around singularities with respect to the original singular metric, however, by replacing $\tilde h$ with $\frac{2w}{\chi(X)}\log G$ near the singularities, the result $h$ solves the Maxwell equation on whole of $X$.
}
Since {the zero mode wavefunction} is expressed as a holomorphic section $u$ of $L\rightarrow X$ (or its complex conjugate), its norm $\braket{u|u}_{\rm hol}$ is locally expressed as
\begin{gather}
	\braket{u|u}_{\rm hol}=\int h |u|^2\,{\rm vol}(X)=\int G^{\frac{2w}{\chi(X)}+2}|u|^2\,dzd\bar{z}.
\end{gather}
This norm must be finite for a normalizable wavefunction.
If $G$ is smooth at $z=0$, the above normalizability condition claims that $u$ does not have a pole at $z=0$.
If there is a brane {with tension $\alpha$} at $z=0$, the above normalizability {demands}
\begin{gather}
	\int_{U_R}G^{\frac{2w}{\chi(X)}+2}|u|^2\,dzd\bar{z}=\int_0^R rdr\int_0^{2\pi}d\theta\,H^{2\alpha\left(\frac{w}{\chi(X)}+1\right)}r^{-2\alpha\left(\frac{w}{\chi(X)}+1\right)}|u|^2<\infty,
\end{gather}
where $U_R=\acom{z\in \mb{C}\,|\,\,|z|<R}$ is an open set around $z=0$ and $H$ is defined by $G=Hr^{-\alpha}$.
Since $H$ is continuous and $H(0)\neq 0$ as we mentioned before, the finiteness of the above integration is equivalent to
\begin{gather}
	\int_0^R rdr\int_0^{2\pi}d\theta\,r^{-2\alpha\left(\frac{w}{\chi(X)}+1\right)}|u|^2<\infty.
\end{gather}
When the Laurent expansion of $u$ around $z=0$ starts from $z^n$, this condition means that $n$ must satisfy
\begin{gather}
\label{NORMALIZABLE}
	n\geq \fl{\alpha\left(\frac{w}{\chi(X)}+1\right)},
\end{gather}
where $\fl{t}$ is an integer part in a real number $t$.

Let us consider $N$ branes on $z_a\in X\,\,(a=1,\dots,N)$ {with brane tensions} $\alpha_a$ respectively.
We define a divisor $D_{{\rm br},\,w}$ by
\begin{gather}
	D_{{\rm br},\,w}=\sum_{a=1}^N\fl{\alpha_a\left(\frac{w}{\chi(X)}+1\right)}z_a.
\end{gather}
Eq.\,\eqref{NORMALIZABLE} implies that a holomorphic section $u$ of $L\rightarrow X$ is normalizable if and only if its divisor $D_u$ satisfies
\begin{gather}
	D_u\geq D_{{\rm br},\,w}.
\end{gather}
Therefore the vector space of normalizable wavefunctions is isomorphic to
\begin{gather}
	\left\{f\neq0:\mbox{meromorphic function}\,|\,{\rm div}\,f+D_L-D_{{\rm br},\, w}\geq0\right\}\cup\{0\},
\end{gather}
where ${\rm div}\,f$ is a divisor defined by poles of a meromorphic function $f$, and $D_L$ is the divisor of $L\rightarrow X$.
The Riemann--Roch theorem counts the dimension $\ell(D_L-D_{{\rm br},w})$ of this vector space:
\begin{gather}
	\ell(D_L-D_{{\rm br},\,w})=w+\frac{1}{2}\chi(X)-\sum_{a=1}^N\fl{\alpha_a\left(\frac{w}{\chi(X)}+1\right)}-\ell(D_{T^*X}-D_L+D_{{\rm br},\,w}),
\end{gather}
{where} $D_{T^*X}$ is a canonical divisor of $X$.
Although it is difficult to compute the last term in many cases, it is neglectable when $X=S^2$ and the degree of $D_L-D_{{\rm br},\,w}$ is non-negative {due to the negativity of ${\rm deg}D_{T^*X}=-\chi(S^2)=-2$}.
Then, we reach the normalizable mode counting formula on $S^2$ in the presence of branes\,:
\begin{gather}
\label{COUNTING}
	\ell(D_L-D_{{\rm br},\,w})=w+1-\sum_{a=1}^N\fl{\alpha_a\left(\frac{w}{2}+1\right)}.
\end{gather}
Note that $w$ is the total Chern number of $L\rightarrow X$.
{In other words}, $w$ in this appendix is a sum of the spin along the extra dimensions and $U(1)$ charge.

\section{Symmetry of magnetized $S^2$}
In general, it is not easy to identify a symmetry which magnetized extra dimensions possess as an isometry.
For example, the 2d torus has $U(1)^2$ {isometry} group {obviously}, however, this is broken once magnetic flux is turned on.
{In considering} a symmetry of the magnetized extra dimensions, we have to {correctly} treat a gauge sector.
The following concept {would be} good guidance for clarifying the symmetry on {magnetized extra dimensions}.

\begin{Def*}
Let $X$ be a complex manifold with a Riemannian metric and $E\rightarrow X$ be a holomorphic vector bundle with a hermitian metric.
	When an automorphism $\varphi:E\rightarrow E$ leaves the hermitian metric invariant and an induced map $X\rightarrow X$ {does} the metric invariant, we {call $\varphi$ an isometry of $E$}.
\end{Def*}

{Indeed}, when {the} magnetic flux is realized as the Chern connection of $E\rightarrow X$, an isometry of $E$ leaves the Chern connection invariant.
{Namely}, the isometry of $E$ does not change a background vector potential, and also by definition an isometry of $E$ does not change the metric of $X$.
Hence, the total isometry of $E$ {reduces to} a symmetry {that the magnetized extra dimensional models themselves possess.}
For example, the breaking of isometries of $T^2$ by magnetic flux can result from the non-existence of such an automorphism \cite{brion2013}.

In this appendix, we realize $S^2$ by a complex projective space $\mb{CP}^1$ because it is useful to understand a global structure of $S^2$.
A projective space $\mb{CP}^1$ is a quotient $(\mb{C}^2 \setminus \{0\})/\sim$, where the equivalent relation $\sim$ is defined by
\begin{gather}
	(x,y)\sim (\lambda x,\lambda y) \qquad (x,y,\lambda\in \mb{C},\,\lambda\neq0).
\end{gather}
{In what follows}, $[x:y]$ denotes the equivalence class of $(x,y)$.
Any abelian magnetic fluxes are realized as holomorphic line bundles on $\mb{CP}^1$ and they are tensor products of a universal line bundle $L^{-1}\rightarrow \mb{CP}^1$,
\begin{gather}
	L^{-1}=\bigcup_{[x:y]\in \mb{CP}^1}{\rm span}_\mb{C}\{(x,y)\}\subseteq \mb{CP}^1\times \mb{C}^2.
\end{gather}
Since the Chern number of $L^{-1}$ is $-1$, that of $L^{qM}:=(L^{-1})^{\otimes(-qM)}$ is $qM\in\mb{Z}$.
A canonical K\"{a}hler potential
\begin{gather}\label{CANONICAL}
	K=\log(|x|^2+|y|^2)
\end{gather}
is invariant under this $SU(2)$, and hence $SU(2)$ is an isometry on the pure sphere.
When two branes are introduced, {the} K\"{a}hler potential is modified to
\begin{gather}\label{TWOBR}
	K=\log(|x|^2+|y|^{2-2\alpha}|x|^{2\alpha}),
\end{gather}
and the original $SU(2)$ isometry on the pure sphere is broken to diagonal $U(1)$.
We define a hermitian metric on $L^{-1}$ as
\begin{gather}\label{HERMIT}
	(\lambda \vec{z},\lambda' \vec{z})\mapsto \bar{\lambda}\lambda' e^{K(x,y)},
\end{gather}
where $\vec{z}=(x,y)$.
By this definition, a field strength of its Chern connection is proportional to the K\"{a}hler form and solves the Maxwell equation.
In a physical language, the Chern connection is exactly a flux on $\mathbb{CP} \simeq S^2$.
When zero or two branes are introduced, we have the K\"{a}hler potential (\ref{CANONICAL}) or (\ref{TWOBR}) which is invariant under the isometry $U(1)$.
When more branes {than three} are introduced, we have no continuous isometry group and only a finite subgroup of $SU(2)$ is realizable.
Since the isometry group is finite, we can find an invariant K\"{a}hler potential.
For {the} invariant K\"{a}hler potential, the hermitian metric (\ref{HERMIT}) on $L^{-1}\rightarrow \mb{CP}^1$ is manifestly invariant under the isometry.\,\footnote{
	We do not have to prepare such an invariant K\"{a}hler potential, actually.
	In fact, when the K\"{a}hler potential is invariant up to (global) K\"{a}hler transformation by modifying the transformation of $L^{-1}$, we can make the hermitian metric given in Eq.\,\eqref{HERMIT} invariant.
}
It is concluded that an isometry is not only a symmetry of the metric, but also of the gauge sector $L^{-1}$, or more general  $L^{qM}$ for an arbitrary integer $qM\in \mb{Z}$.

\bibliographystyle{utphys}
\bibliography{references}
\end{document}